# Impact of self-association on the architectural properties of bacterial nucleoid proteins

Marc Joyeux[(#)]
*Laboratoire Interdisciplinaire de Physique,
CNRS and Université Grenoble Alpes, Grenoble, France*

**Running title**: Self-association of nucleoid proteins.

**Abstract:** The chromosomal DNA of bacteria is folded into a compact body called the nucleoid, which is composed essentially of DNA (≈80%), RNA (≈10%), and a number of different proteins (≈10%). These nucleoid proteins act as regulators of gene expression and influence the organization of the nucleoid by bridging, bending, or wrapping the DNA. These so-called architectural properties of nucleoid proteins are still poorly understood. For example, the reason why certain proteins compact the DNA coil in certain environments but make instead the DNA more rigid in other environments is the matter of ongoing debates. Here, we address the question of the impact of the self-association of nucleoid proteins on their architectural properties and try to determine whether differences in self-association are sufficient to induce large changes in the organization of the DNA coil. More specifically, we developed two coarse-grained models of proteins, which interact identically with the DNA but self-associate differently by forming either clusters or filaments in the absence of the DNA. We showed through Brownian dynamics simulations that self-association of the proteins increases dramatically their ability to shape the DNA coil. Moreover, we observed that cluster-forming proteins compact significantly the DNA coil (similar to the DNA-bridging mode of H-NS proteins), whereas filament-forming proteins increase instead significantly the stiffness of the DNA chain (similar to the DNA-stiffening mode of H-NS proteins). This work consequently suggests that the knowledge of the DNA-binding properties of the proteins is in itself not sufficient to understand their architectural properties. Rather, their self-association properties must also be investigated in detail, because they might actually drive the formation of different DNA/protein complexes.

(#) email : marc.joyeux@univ-grenoble-alpes.fr

**Statement of significance:** Many nucleoid proteins have two interrelated functions: They act as regulators of gene expression and shape the nucleoid by bridging, bending or wrapping the DNA. It is usually accepted that the way these proteins bind to the DNA dictates the way they shape the DNA coil. For example, proteins that bridge distal DNA segments are expected to compact the nucleoid. Through coarse-grained modeling and Brownian dynamics simulations, we identify here yet another key parameter and show that protein self-association impacts very profoundly their architectural properties. Two proteins, which interact similarly with the DNA but oligomerize differently, may have strikingly different architectural properties, with one protein compacting the DNA coil and the other one making instead the DNA molecule more rigid.

# INTRODUCTION

Bacteria lack a nucleus, but their chromosomal DNA is nevertheless folded into a compact body called the nucleoid, which is markedly different from the rest of the cytoplasm. The nucleoid is composed essentially of DNA ($\approx$80%), RNA ($\approx$10%), and a number of different proteins ($\approx$10%) (1,2). These proteins act as regulators of gene expression (3-5) and influence the organization of the nucleoid by bridging, bending, or wrapping the DNA (5-8). There are at least 12 different species of nucleoid proteins (9), among which HU (10), IHF (11), H-NS (12), Fis (13) and Lrp (14) have been extensively studied. It has been shown that the abundance of many of the nucleoid proteins varies dramatically in response to changes in the growth rate of the cell (15). Their occupancy landscape in the nucleoid (16) and along the genome (17) has also been investigated.

The mechanisms by which nucleoid proteins shape the DNA are still poorly understood. This is due, in part, to the fact that architectural properties are specific to each protein. For example, H-NS, ParB and SMC form bridges between two DNA segments, but these bridges are qualitatively different and affect chromosome organization and gene regulation in contrasting ways (18). Moreover, several proteins exhibit dual architectural properties, depending on several factors, like the concentration of proteins and the DNA binding sequence (8). For example, HU is essentially known for its DNA-bending capabilities and Lrp for its DNA-bridging capabilities, but both of them are also able to wrap the DNA (8). Finally, subtle variations of the cytosol may alter dramatically the architectural properties of certain proteins. For example, an increase in the concentration of divalent cations in the cytosol causes H-NS to switch from the DNA-stiffening mode (characterized by rigid DNA/H-NS complexes) to the DNA-bridging mode (characterized by more compact DNA coils) (19).

Through the development of coarse-grained models and Brownian dynamics simulations, we recently showed that the switch of H-NS proteins from the DNA-stiffening to the DNA-bridging mode may be due to the fact that an increase in the concentration of multivalent cations provokes an increase in the screening of electrostatic charges along the DNA backbone, which leads in turn to a decrease in the strength of DNA/protein interactions compared to protein/protein interactions (20). As a consequence, for concentrations of multivalent cations smaller than a certain threshold, proteins form filaments which stretch along the DNA molecule. In contrast, for larger concentrations of multivalent cations, proteins form clusters which connect genomically distant DNA sites (20). We argued that these two types of DNA/protein complexes may correspond to the DNA-stiffening and DNA-bridging modes of H-NS, respectively. Unfortunately, the model was not precise enough for protein filaments to increase the effective stiffness of the DNA chain and for protein clusters to reduce significantly the radius of the DNA coil. Moreover, this first study left an important question unanswered, namely: *To what extent do the self-association properties of proteins influence their nucleoid architectural properties* ? In the model proposed in (20), proteins self-associate in the form of 3-dimensional clusters and the final conformation of DNA/protein complexes is actually driven by the relative strength of DNA/protein interactions compared to protein/protein ones. The question we address in the present work is different, in the sense that we consider two proteins, which interact identically with the DNA chain but self-associate differently, and we want to determine whether such a difference in self-association properties is sufficient to induce large changes in the organization of the DNA coil. In addition to its obvious relevance in prokaryotes, this question may also be of fundamental importance in eukaryotes, since it has recently been shown that slightly different variants of histone proteins form central tetramers with rather different properties, which may potentially influence drastically nucleosome assembly and disassembly (21). In order to answer this question, we developed two new coarse-grained models of proteins, which interact identically with the DNA chain but self-associate differently. In the absence of DNA, Model I proteins form spontaneously clusters, whereas Model II proteins form filaments. We showed through Brownian dynamics simulations that self-association of the protein chains increases dramatically their ability to shape the DNA coil. Moreover, we observed that Model I proteins compact significantly the DNA coil (similar to the DNA-bridging mode of H-NS), whereas Model II proteins increase instead significantly the stiffness of the DNA chain (similar to the DNA-stiffening mode of H-NS). This work consequently suggests that the knowledge of the DNA-binding properties of given proteins is in itself not sufficient to understand their architectural properties. Rather, their self-association properties must also be investigated in detail, because they might actually drive the formation of different DNA/protein complexes.

**METHODS**

    **Model**

The two coarse-grained bead-and-spring models developed in the course of the present study are described in detail in Model and Simulations in Supporting Material. In brief, they consist of a long DNA chain and 200 short protein chains enclosed in a confinement sphere. Each chain is composed of beads of equal size connected by springs. Concentrations of nucleotides and proteins are of the same order of magnitude as *in vivo* ones. For DNA, each bead represents 7.5 base pairs (bp) and the chain contains 2880 beads, equivalent to 21600 bp, as in (20,22,23). Each protein chain contains 7 beads with index $m$ ($1 \leq m \leq 7$), where terminal beads $m=1$ and 7 represent the two DNA-binding sites of each protein, whereas beads $m=2$ and 6 (for Model I) or $m=2$, 3, 5 and 6 (for Model II) represent the isomerization sites of the protein (Fig. 1). The overall potential energy of the system (Eq. (S24)) is the sum of four terms, which describe the internal energy of the DNA chain (Eq. (S1)), the internal energy of protein chains (Eq. (S8)), DNA/protein interactions (Eq. (S14)), and protein/protein interactions (Eq. (S18)). The first three terms are common to Model I and Model II. In particular, for both models, the two terminal beads of each protein chain ($m=1$ and $m=7$) can rotate without energy penalty around beads $m=2$ and $m=6$, respectively (Eq. (S10)), and can bind to the DNA chain with a maximum binding energy of $-7.8\, k_B T$ (Eqs. (S15) and (S16), Fig. S2(a)). Because of the free rotation of terminal beads, protein chains are significantly less rigid than the DNA chain, as is usually the case *in vivo*. Moreover, the DNA/protein binding energy is comparable to experimentally determined values for complexes of DNA and H-NS ($\approx -11.0\, k_B T$ (24)). Model I and II protein chains differ only in their isomerization properties. Indeed, for Model I, beads $m=2$ and $m=6$ of one protein chain may bind to beads $m=2$ and $m=6$ of other protein chains, whereas, for Model II, beads $m=2$ and $m=6$ of one protein chain may bind to beads $m=3$ and $m=5$ of other protein chains (Eqs. (S19), (S21) and (S23), Figs. 1 and S2(b)). As a result, Model I protein chains form spontaneously clusters, whereas Model II protein chains form filaments. The binding interaction between two protein isomerization beads is modeled by a Lennard-Jones 3-6 potential of depth $\varepsilon_{LJ}$ (Eq. (S19)). The isomerization binding energy for Model I is $-\varepsilon_{LJ}$, whereas it varies with a slope close to $-2\varepsilon_{LJ}$ for Model II (Fig. S3). For comparison, it is reminded that the experimentally determined value of the enthalpy change upon forming a complex between two H-NS dimers is $-10.2\, k_B T$ (25).

**Simulations**

The dynamics of the models was investigated by integrating numerically Langevin equations of motion with kinetic energy terms dropped and time steps of 1.0 ps. Temperature $T$ was assumed to be 298 K throughout the study. The value of the Debye length used in the simulations ($r_D = 1.07$ nm) corresponds to a concentration of monovalent salt of 100 mM, which is the value that is generally assumed for the cytoplasm of bacterial cells. After each integration step, the position of the centre of the confining sphere was slightly adjusted so as to coincide with the centre of mass of the DNA molecule, in order that compact DNA/protein complexes do not stick to the wall of the confinement sphere and results are as little as possible affected by the interactions with the wall (26). Simulations were run for both models and values of $\varepsilon_{LJ}$ ranging from $4 k_B T$ to $12 k_B T$, in order to check the impact of self-association of the protein chains on the equilibrium properties of the system. The upper limit was fixed to $12 k_B T$, because the probability for Model II proteins to form clusters instead of filaments becomes non negligible for this value of $\varepsilon_{LJ}$ and increases rapidly for larger values.

**RESULTS AND DISCUSSION**

**Self-association of Model I and II protein chains.**

Model I and II protein chains interact identically with the DNA chain but self-associate differently. The goal of this work is to determine whether the difference in self-association might result in different architectural properties of the proteins, that is, in DNA/protein complexes with substantially different conformations. A preliminary step consists in characterizing in some detail the complexes that protein chains form spontaneously in the absence of the DNA chain. To this end, 200 protein chains were introduced at random non-overlapping positions in the confinement sphere and the system was allowed to equilibrate for values of $\varepsilon_{LJ}$ (the depth of the Lennard-Jones 3-6 potential that governs protein/protein interactions) ranging from $4 k_B T$ to $12 k_B T$. Typical equilibration times range from 1 ms to 50 ms, depending on the model and the value of $\varepsilon_{LJ}$.

For the lowest values of $\varepsilon_{LJ}$, thermal noise is strong enough to prevent association of the protein chains, as was also the case for a previous model of H-NS mediated compaction of bacterial DNA (27,28). In contrast, for larger values of $\varepsilon_{LJ}$, Model I proteins form clusters, whereas Model II proteins form filaments. Representative snapshots of equilibrated conformations are shown in the top row of Fig. 2. Evolution of protein complexes with increasing values of $\varepsilon_{LJ}$ may be characterized by plotting $q(s)$, the probability distribution for a protein chain to bind to $s$ other protein chains. For this purpose, it was considered that two protein chains are bound if the interaction between at least two of their isomerization beads is attractive and of magnitude larger than $3\,k_B T$. The choice of the $3\,k_B T$ threshold is somewhat arbitrary, but the principal features of the distributions shown in Fig. S4 do not depend critically thereon. For Model I (left column of Fig. S4), protein chains do not associate significantly up to $\varepsilon_{LJ} = 6\,k_B T$, whereas for $\varepsilon_{LJ} \geq 7\,k_B T$ each protein chain binds on average to 4 or 5 other protein chains, which results in clusters like the ones shown in the top left vignette of Fig. 2. For Model II (right column of Fig. S4), protein chains do not associate significantly up to $\varepsilon_{LJ} = 8\,k_B T$, whereas for $\varepsilon_{LJ} \geq 9\,k_B T$ each protein chain binds at maximum to 2 other protein chains, which results in filaments like the ones shown in the top right vignette of Fig. 2.

**Complexes of DNA and protein chains.**

Let us now consider complexes formed by the DNA chain and Model I and II protein chains. These complexes were obtained by first allowing the DNA chain to equilibrate inside the confinement sphere. The 200 protein chains were then introduced at random non-overlapping positions in the confinement sphere and the system was allowed to equilibrate again for values of $\varepsilon_{LJ}$ ranging from $4\,k_B T$ to $12\,k_B T$. Typical equilibration times range from 1 ms to 20 ms, depending on the model and the value of $\varepsilon_{LJ}$. For the lowest values of $\varepsilon_{LJ}$, equilibrated conformations display few protein/protein contacts and a limited number of DNA/protein contacts, whereas huge DNA/protein complexes are observed for larger values of $\varepsilon_{LJ}$. Representative snapshots of equilibrated conformations obtained with large values of $\varepsilon_{LJ}$ are shown in the bottom row of Fig. 2. For Model I (bottom left vignette of Fig. 2), the DNA chain wraps around the protein clusters, which are quite similar to those obtained without the DNA. In contrast, for Model II (bottom right vignette of Fig. 2), the protein filaments and the DNA chain form thick bundles, in which they align parallel to each other.

As schematized in Fig. S5, protein chains in thermodynamic equilibrium with a DNA chain can be described either as free (no contact with the DNA chain), dangling (only one extremity of the protein chain binds to the DNA chain), *cis*-bound (the two extremities of the protein chain bind to genomically close DNA beads), or bridging (the two extremities of the protein chain bind to genomically distant DNA beads). The evolution with increasing values of $\varepsilon_{LJ}$ of the average fraction of the four types of protein chains is shown in Fig. 3 for Model I (open symbols) and II (filled symbols). Because the two models of protein interact similarly with the DNA chain, the curves for Model I and II remain superposed as long as self-association of protein chains remains negligible, that is, up to $\varepsilon_{LJ} = 6\,k_BT$. In this regime, ≈50% of the protein chains are free, ≈22% are dangling, ≈22% are *cis*-bound, and only ≈6% are bridging the DNA chain. However, the onset of protein self-association is accompanied in both models by a strong decrease in the number of free proteins (≈10% at $\varepsilon_{LJ} = 12\,k_BT$), which is compensated by a strong increase in the number of bridging proteins (≈35% at $\varepsilon_{LJ} = 12\,k_BT$). Evolution with increasing values of $\varepsilon_{LJ}$ is sharper for Model I than for Model II.

The two models differ in that the fraction of *cis*-bound protein chains increases up to ≈35% at $\varepsilon_{LJ} = 12\,k_BT$ for Model II, whereas it remains nearly constant at ≈25% for Model I. However, this discrepancy merely reflects different organizations of DNA/protein clusters, because the plot, as a function of $\varepsilon_{LJ}$, of the average fraction of protein chains which belong to clusters that bridge genomically distant DNA beads indicates that the onset of protein self-association is accompanied in both models by an increase in the fraction of such protein chains from about 6% to 100% (see Fig. S6).

Finally, it may be worth noting that the maximum number of bridging proteins observed for Model I and $\varepsilon_{LJ} = 7\,k_BT$ (≈42%) is due to the fact that, for this value of $\varepsilon_{LJ}$, which is the smallest one that leads to protein self-association for Model I, the protein chains still display some ability to escape and rearrange after binding to an existing assembly. As a result, for $\varepsilon_{LJ} = 7\,k_BT$ protein chains form a single regular (nearly 2D) sheet, which apparently maximizes the number of bridges, whereas they assemble in more irregular 3D clusters for larger values of $\varepsilon_{LJ}$. This can be checked in Fig. S7, which shows typical conformations obtained for Model I and $\varepsilon_{LJ} = 6\,k_BT$, $7\,k_BT$, and $8\,k_BT$. For Model I and $\varepsilon_{LJ} \geq 8\,k_BT$, the number of bridges remains constant within computational uncertainties.

The plots of the probability distribution for a protein chain to bind to *s* other protein chains, $q(s)$, are shown in Fig. S8 for equilibrated DNA/protein systems. These plots show that the presence of the DNA chain does not alter significantly the self-association of protein chains. Indeed, Model I protein chains bind on average to 4 or 5 other protein chains starting from $\varepsilon_{LJ} = 7\,k_B T$ (left column of Fig. S8), whereas Model II protein chains bind at maximum to 2 other protein chains starting from $\varepsilon_{LJ} = 9\,k_B T$ (right column of Fig. S8), as is also the case without the DNA (Fig. S4). The plots of $p(s)$, the probability distribution for a DNA-binding protein bead to bind to $s$ DNA beads, are also shown in Fig. S8. As for $q(s)$, it was considered that a protein bead and a DNA bead are bound if their interaction is attractive and of magnitude larger than $3\,k_B T$. A first conclusion concerning the impact of the self-association of protein chains on DNA/protein complexes can be drawn from the comparison of the plots of $p(s)$ and $q(s)$ in Fig. S8. This figure reveals that the binding of protein chains to the DNA chain is boosted by protein self-association. Indeed, for both models, only ≈30% of the DNA-binding protein beads bind to a DNA bead ($p(0) \approx 0.7$) for values of $\varepsilon_{LJ}$ corresponding to weak protein self-association, that is, up to $\varepsilon_{LJ} = 6\,k_B T$ for Model I and $\varepsilon_{LJ} = 8\,k_B T$ for Model II. In contrast, ≈70% of the DNA-binding protein beads bind to at least one DNA bead ($p(0) \approx 0.3$) for larger values of $\varepsilon_{LJ}$, for which strong protein self-association is observed.

**Architectural properties of Model I and II protein chains differ widely.**

Let us now examine in more detail the extent to which the self-association of Model I and II protein chains impact their architectural properties. To this end, we studied the evolution of two quantities which describe the geometrical properties of the DNA, namely the mean radius of the coil, $\langle R \rangle$, and the persistence length of the DNA chain, $\xi$.

The mean radius of the DNA coil, $\langle R \rangle$, is defined according to

$$\langle R \rangle = \frac{1}{n} \sum_{k=1}^{n} \| \mathbf{r}_k - \mathbf{r}_{CM} \|, \qquad (1)$$

where $\mathbf{r}_k$ denotes the position of the center of DNA bead $k$ and $\mathbf{r}_{CM}$ the position of the center of mass of the DNA coil. In the absence of protein chains, the average value $\langle R \rangle = 82.1\,\text{nm}$ results from the balance of the compressive forces exerted by the confinement sphere and the expansive forces arising from the bending rigidity of the DNA chain and the electrostatic repulsion between DNA beads. As long as protein chains self-associate only weakly, addition of 200 of them inside the confinement sphere has little effect on the mean radius of the DNA coil, as can be checked in the top plot of Fig. 4 which shows the evolution of $\langle R \rangle$ with increasing values of $\varepsilon_{\text{LJ}}$. Indeed, $\langle R \rangle$ remains close to 80 nm for small values of $\varepsilon_{\text{LJ}}$. This is a direct consequence of the fact that only ≈6% of the proteins bridge the DNA chain, which is insufficient to compact significantly the DNA coil. In contrast, $\langle R \rangle$ decreases rapidly below 70 nm for Model I and values of $\varepsilon_{\text{LJ}}$ larger than $7\,k_{\text{B}}T$. This indicates that the ≈35% of protein chains, which bridge the DNA chain (bottom left vignette of Fig. 2), are quite efficient in compacting the DNA coil. However, this is not the case for Model II and $\varepsilon_{\text{LJ}} > 9\,k_{\text{B}}T$, although approximately 35% of the protein chains also bridge the DNA chain. The reason is that most of these bridges localize in thick bundles similar to the one shown in the bottom right vignette of Fig. 2, where they essentially work to maintain DNA segments parallel to each other. Such distributions of bridges are apparently not as efficient for compacting the DNA coil as the distributions of bridges for Model I proteins. An exception occurs for Model II and $\varepsilon_{\text{LJ}} = 9\,k_{\text{B}}T$, for which significant compaction of the DNA coil ($\langle R \rangle \approx 73$ nm) is observed in the top plot of Fig. 4. The reason is that this value of $\varepsilon_{\text{LJ}}$ is the smallest one that leads to protein self-association for Model II, and DNA/protein complexes are different from the thick bundles observed for larger values of $\varepsilon_{\text{LJ}}$. They resemble more the conformations obtained with Model I, as can be checked in Fig. S9, which shows typical conformations obtained for Model II and $\varepsilon_{\text{LJ}} = 8\,k_{\text{B}}T$, $9\,k_{\text{B}}T$, and $10\,k_{\text{B}}T$.

Conclusion is therefore that neither Model I nor Model II protein chains are efficient in compacting the DNA coil when in the monomer form. In contrast, Model I protein chains compact significantly the DNA coil as soon as they self-associate (a similar result was already obtained with a different model (29)). This is not the case for Model II protein chains (except for $\varepsilon_{\text{LJ}} = 9\,k_{\text{B}}T$), in spite of the fact that all protein chains interact similarly with the DNA chain.

Let us now consider the persistence length of the DNA chain, $\xi$, which is related to the directional correlation function $C(\Delta) = \langle \mathbf{t}(x).\mathbf{t}(x+\Delta) \rangle$, where $\mathbf{t}(x)$ denotes the unit vector tangent to the DNA chain at curvilinear position $x$, according to

$$C(\Delta) = \exp(-\frac{\Delta}{\xi}) \,. \qquad (2)$$

Practically, the directional correlation function can be estimated from

$$C(Ll_0) \approx \frac{1}{nN} \sum_{c=1}^{N} \sum_{k=1}^{n} \mathbf{t}_{c,k} \cdot \mathbf{t}_{c,k+L} , \qquad (3)$$

where $N$ is a large number of DNA conformations spanning a large time interval, and $\mathbf{t}_{c,k}$ is the unit vector tangent to the DNA chain at the center of bead $k$ in conformation $c$. The persistence length $\xi$ is obtained from an exponential fit of the evolution of $C(Ll_0)$ as a function of $L$ over a certain interval of values of $L$. By using this procedure for $0 \leq L \leq 20$, we obtained $\xi = 41.7$ nm for the DNA chain enclosed in the confinement sphere without protein chains. This value is somewhat smaller than the value estimated from the bending rigidity of the DNA chain ($\xi = 50$ nm). This is due to the fact that the confinement sphere imposes non-negligible additional curvature to the DNA chain, because its diameter ($2R_0 = 240$ nm) is only $\approx 5$ times larger than the persistence length of unconstrained DNA. For equilibrated DNA/protein complexes, it is interesting to discriminate between the persistence length of DNA segments which are not bound to any protein chain ($\xi_{\text{free}}$) and the persistence length of DNA segments bound to at least one protein chain ($\xi_{\text{bound}}$). This is easily achieved by testing at each step of the averaging procedure whether any bead of the DNA segment comprised between beads $k$ and $k+L$ binds to a protein chain or not, and using this segment adequately to compute either $\xi_{\text{free}}$ or $\xi_{\text{bound}}$.

The evolution of $\xi_{\text{free}}$ and $\xi_{\text{bound}}$ with increasing values of $\varepsilon_{\text{LJ}}$ is shown in the bottom plot of Fig. 4. Not surprisingly, $\xi_{\text{free}}$ remains close to 40 nm for all values of $\varepsilon_{\text{LJ}}$. Moreover, addition of 200 protein chains inside the confinement sphere has little effect on $\xi_{\text{bound}}$ as long as protein chains do not self-associate significantly. This indicates that the $\approx 22\%$ of protein chains which bind to the DNA chain in *cis* do not increase significantly its rigidity. In contrast, $\xi_{\text{bound}}$ increases rapidly up to $\approx 80$ nm for Model II and values of $\varepsilon_{\text{LJ}}$ larger than $10 \, k_B T$. This confirms that the thick bundles composed of DNA segments maintained parallel to each other and bridged by protein segments are quite rigid, as could be anticipated from their almost rectilinear shape (bottom right vignette of Fig. 2). No increase in $\xi_{\text{bound}}$ is however observed for Model I, even for large values of $\varepsilon_{\text{LJ}}$, as can be checked in Fig. 4. Although the network of protein chains formed for $\varepsilon_{\text{LJ}} \geq 7 \, k_B T$ is quite efficient in compacting the DNA coil, it is flexible enough for the numerous cross-links not to alter significantly the persistence length of DNA segments bound to protein chains.

Conclusion is consequently that neither Model I nor Model II protein chains are efficient in altering the persistence length of the DNA coil when in the monomer form. In contrast, when Model II protein chains self-associate, the persistence length of DNA segments localized in the thick bundles formed by DNA/protein complexes is twice as large as that of free DNA segments. This is however not the case for Model I protein chains, in spite of the fact that all protein chains interact similarly with the DNA chain.

**Discussion**

In this work, we studied the properties of two models describing non-specific interactions between circular DNA and nucleoid proteins. The DNA/protein interaction potential is the same for the two models and was kept constant in all simulations. In contrast, when the strength of protein/protein interactions is large enough, Model I proteins self-associate in the form of clusters, whereas Model II proteins form filaments. The strength of protein/protein interactions was varied systematically in the simulations, in order to check the impact of protein self-association on the geometrical and mechanical properties of DNA/protein complexes. The two models display characteristic features:

- for the two models, binding of the proteins to the DNA increases strongly when proteins self-associate, although the strength of DNA/protein interactions is kept constant,
- when in the monomer form, neither Model I nor Model II proteins are efficient in compacting the DNA coil or increasing the rigidity of the DNA,
- clusters of Model I proteins compact significantly the DNA coil, but this is not the case for filaments of Model II proteins, although all proteins interact similarly with the DNA,
- filaments of Model II proteins increase significantly the rigidity of the DNA, but this is not the case for clusters of Model I proteins, although all proteins interact similarly with the DNA.

These models consequently suggest that the self-association of nucleoid proteins may have a rich and profound impact on their architectural properties. This claim and the models proposed here are supported by a set of experimental results:

First, many of the nucleoid proteins can self-associate and are present in cells in polymeric forms. For example, H-NS proteins form dimers at low concentrations but assemble into larger multimers at higher concentrations (25,30,31). Other members of the H-NS family, like StpA, can also self-associate (31,32). As for the models proposed here, cooperative binding of H-NS to DNA is related to protein-protein interactions (33). The resulting filaments of H-NS proteins bound to the DNA substrate are clearly seen in crystallographic experiments (34). It is believed that such protein filaments block DNA accessibility and are the structural basis for gene silencing (35,36), which is one of the main roles of H-NS in the cells. It has however been shown that the simple coverage of the DNA substrate by H-NS proteins at high concentrations is not sufficient, and that the capacity of proteins to self-associate is crucial for the regulation of gene expression: Derivatives of H-NS that are unable to oligomerize fail in silencing genes (37-40).

Moreover, two nucleoid proteins present in the stationary phase, Dps and CbpA, can also self-associate and experiments have shown that their aggregation and the compaction of the DNA are parallel phenomena (41,42). Unlike H-NS (26), Dps molecules do not align in filaments in co-crytals, but are rather packed in pseudohexagonal layers (43). The layers slide along the DNA direction and enable the formation of grooves for DNA accommodation (43). Similarly, partition proteins ParB *in vivo* first bind to the specific *parS* site and then spread, that is, they simultaneously self-assemble stochastically and bind to the DNA away from the *parS* site, thus bridging the DNA (44-46). Spreading ability is required, as ParB mutants which lack this ability are also defective in partition (47).

According to this short digest of experimental results, it appears that Model I captures adequately the main features of Dps, CbpA and ParB, which must assemble in clusters to bind to and compact the DNA molecule. The case of H-NS is more complex. Owing to the crystallographic structure in (34), which displays H-NS filaments aligned parallel to the DNA molecule, as well as the experimental observation that the persistence length of DNA/H-NS complexes may be as large as 130 nm at low divalent cations concentration (48), it is tempting to conclude that Model II provides a correct description of DNA/H-NS interactions in this salt regime. However, the fact that H-NS proteins form filaments when bound to the DNA is not a proof that they also do so when the substrate is lacking (20). The switch from the DNA-stiffening mode to the DNA-bridging mode of H-NS at higher divalent cations concentrations (19) may also be tentatively interpreted as an indication that the self-association properties of H-NS switch from Model II-type to Model I-type. In this respect, we note with interest that it has recently been shown that environmental variations have a direct effect on the self-association properties of H-NS (49). More work is however clearly needed to ascertain whether the switch from the DNA-stiffening mode to the DNA-bridging mode of H-NS is due to a decrease in the strength of DNA/protein interactions, as proposed in (20), a variation in the self-association mode of proteins, as suggested by the present work, or variations in the geometry of the H-NS molecule (50).

**CONCLUSION**

In this work, we used coarse-grained modeling to investigate the impact of the self-association of nucleoid proteins on their architectural properties. The simulations suggest that this impact is probably strong and that different modes of self-association may result in different architectural capabilities of the proteins. Self-association is therefore a property of the proteins that is worth considering when trying to understand how they shape the DNA coil.

To conclude, we would like to mention that models similar to those discussed in the present work have recently been proposed to study the formation of the bacterial nucleoid through the demixing of DNA and non-binding globular macromolecules (51-54), the preferential localization of the nucleoid inside the cell (26), the mechanism of facilitated diffusion, by which proteins search for their targets along the DNA sequence (55-57), and the requirements for DNA-bridging proteins to act as topological barriers of the bacterial genome (23). All these models are compatible and it is possible to combine two (or more) of them to get a more complete and realistic description of bacterial cells (22). This point is crucial, since the effects of different processes taking place simultaneously in living cells are not simply additive and the outcome may be difficult to predict when considering only the effects of each mechanism taken separately (22,29). For example, both DNA/macromolecules demixing and DNA supercoiling contribute to the compaction of the bacterial DNA, but the total compaction of the DNA coil is the sum of the two contributions only in a limited range of values of macromolecular concentration and superhelical density, whereas their interplay is much more complex outside from this range (22). In this respect, it will certainly be instructive in future work to use the models discussed in the present work to investigate the interplay of nucleoid proteins and macromolecular crowders (29,52-54), transcription factors (55-57), or DNA supercoiling and topological insulators (23).

**SUPPORTING MATERIAL**
Model and Simulations section. Figures S1 to S9.

**SUPPORTING CITATIONS**

References (58-70) appear in the Supporting Material.

# REFERENCES

1. Stonington O. G., and D. E. Pettijohn. 1971. The folded genome of *Escherichia coli* isolated in a protein-DNA-RNA complex. *Proc. Natl. Acad. Sci. USA*. 68:6-9.

2. Worcel A., and E. Burgi. 1972. On the structure of the folded chromosome of *Escherichia coli*. *J. Mol. Biol.* 71:127-47.

3 Browning D. F., D. C. Grainger, and S. J. W. Busby. 2010. Effects of nucleoid-associated proteins on bacterial chromosome structure and gene expression. *Curr. Opin. Microbiol.* 13:773-780.

4. Dillon S. C., A. D. S. Cameron, K. Hokamp, S. Lucchini, J. C. D. Hinton, and C. J. Dorman. 2010. Genome-wide analysis of the H-NS and Sfh regulatory networks in S*almonella Typhimurium* identifies a plasmid-encoded transcription silencing mechanism. *Mol. Microbiol.* 76:1250-1265.

5. Dillon, S. C., and C. J. Dorman. 2010. Bacterial nucleoid-associated proteins, nucleoid structure and gene expression. *Nat. Rev. Microbiol.* 8:185-195.

6. Johnson, R. C., L. M. Johnson, J. W. Schmidt, and J. F. Gardner. 2005. Major nucleoid proteins in the structure and function of the *Escherichia coli* chromosome. In The bacterial Chromosome. N.P. Higgins, editor. ASM, Washington DC. Pp. 65-132.

7. Dame, R. T. 2005. The role of nucleoid-associated proteins in the organization and compaction of bacterial chromatin. *Mol. Microbiol.* 56:858-870.

8. Luijsterburg, M. S., M. F. White, R. van Driel, and R. T. Dame. 2008. The major architects of chromatin: Architectural proteins in bacteria, archaea and eukaryotes. *Crit. Rev. Biochem. Mol. Biol.* 43:393-418.

9. Azam T. A., and A. Ishihama. 1999. Twelve species of the nucleoid-associated protein from *Escherichia coli*. Sequence recognition specificity and DNA binding affinity. *J. Biol. Chem.* 274:33105–33113.

10. Pinson V., M. Takahashi, and J. Rouviere-Yaniv. 1999. Differential binding of the *Escherichia coli* HU, homodimeric forms and heterodimeric form to linear, gapped and cruciform DNA. *J. Mol. Biol.* 287:485-497.

11. Wang S., R. Cosstick, J. F. Gardner, and R. I. Gumport. 1995. The specific binding of *Escherichia coli* integration host factor involves both major and minor grooves of DNA. *Biochemistry* 34:13082–13090.

12. Gulvady R., Y. Gao, L. J Kenney, and J. Yan. 2018. A single molecule analysis of H-NS uncouples DNA binding affinity from DNA specificity. *Nucleic Acids Res.* 46:10216-10224.

13. Stella S., D. Cascio, and R. C. Johnson. 2010. The shape of the DNA minor groove directs binding by the DNA-bending protein Fis. *Genes Dev.* 24:814-826.

14. Peterson, S. N., F. W. Dahlquist, and N. O. Reich. 2007. The role of high affinity non-specific DNA binding by Lrp in transcriptional regulation and DNA organization. *J. Mol. Biol.* 369:1307-1317.

15. Azam T. A., A. Iwata, A. Nishimura, S. Ueda, and A. Ishihama. 1999. Growth phase-dependent variation in protein composition of the *Escherichia coli* nucleoid. *J. Bacteriol.* 181:6361-6370.

16. Azam, T. A., S. Hiraga, and A. Ishihama. 2000. Two types of localization of the DNA-binding proteins within the *Escherichia coli nucleoid. Genes to Cells.* 5:613-626.

17. Vora T., A. K. Hottes, and S. Tavazoie. 2009. Protein occupancy landscape of a bacterial genome. *Mol. Cell*. 35:247-253.

18. Song, D., and J. J. Loparo. 2015. Building bridges within the bacterial chromosome. *Trends Genet.* 31:164-173.

19. Liu Y. J., H. Chen, L. J. Kenney, and J. Yan. 2010. A divalent switch drives H-NS/DNA-binding conformations between stiffening and bridging modes. *Genes Dev.* 24:339-344.

20. Joyeux, M. 2018. Role of salt valency in the switch of H-NS proteins between DNA-bridging and DNA-stiffening modes. *Biophys. J.* 114:2317-2325.

21. Zhao, H., D. Winogradoff, Y. Dalal, and G. A. Papoian. 2019. The oligomerization landscape of histones. *Biophys. J.* 116:1845-1855.

22. Joyeux, M. 2020. Bacterial nucleoid: Interplay of DNA demixing and supercoiling. *Biophys. J.* 118:2141-2150.

23. Joyeux, M., and I. Junier. 2020. Requirements for DNA-bridging proteins to act as topological barriers of the bacterial genome. *Biophys. J.* 119:1215-1225.

24. Ono, S., M. D. Goldberg, T. Olsson, D. Esposito, J. C. D. Hinton, and J. E. Ladbury. 2005. H-NS is a part of a thermally controlled mechanism for bacterial gene regulation. *Biochem. J.* 391:203-213.

25. Ceschini, S., G. Lupidi, M. Coletta, C. L. Pon, E. Fioretti, and M. Angeletti. 2000. Multimeric self-assembly equilibria involving the histone-like protein H-NS. A thermodynamic study. *J. Biol. Chem.* 275:729-734.

26. Joyeux, M. 2019. Preferential localization of the bacterial nucleoid. *Microorganisms*. 7:204.

27. Joyeux, M., and J. Vreede. 2013. A model of H-NS mediated compaction of bacterial DNA. *Biophys. J.* 104:1615-1622.

28. Joyeux, M. 2014. Equilibration of complexes of DNA and H-NS proteins on charged surfaces: A coarse-grained model point of view. *J. Chem. Phys.* 141:115102.

29. Dias, R. S. 2019. Role of protein self-association on DNA condensation and nucleoid stability in a bacterial cell model. *Polymers*. 11:1102.

30. Smyth, C. P., T. Lundbäck, D. Renzoni, G. Siligardi, R. Beavil, M. Layton, J. M. Sidebotham, J. C. D. Hinton, P. C. Driscoll, C. F. Higgins, and J. E. Ladbury. 2000. Oligomerization of the chromatin-structuring protein H-NS. *Mol. Microbiol.* 36:962–972.

31. Leonard, P. G., S. Ono, J. Gor, S. J. Perkins, and J. E. Ladbury. 2009. Investigation of the self-association and hetero-association interactions of H-NS and StpA from Enterobacteria. *Mol. Microbiol.* 73:165-179.

32. Suzuki, C., K. Kawazuma, S. Horita, T. Terada, M. Tanokura, K. Okada, H. Yamane, and H. Nojiri. 2014. Oligomerization mechanisms of an H-NS family protein, Pmr, encoded on the plasmid pCAR1 provide a molecular basis for functions of H-NS family members. *PLoS ONE* 9:e105656.

33. Giangrossi, M., K.Wintraecken, R. Spurio, and R. de Vries. 2014. Probing the relation between protein–protein interactions and DNA binding for a linker mutant of the bacterial nucleoid protein H-NS. *Biochim. Biophys. Acta* 1844:339–345.

34. Arold, S. T., P. G. Leonard, G. N. Parkinson, and J. E. Ladbury. 2010. H-NS forms a superhelical protein scaffold for DNA condensation. *P. Natl. Acad. Sci. USA*. 107:15728-15732.

35. Lim, C. J., S. Y. Lee, L. J. Kenney, and J. Yan. 2012. Nucleoprotein filament formation is the structural basis for bacterial protein H-NS gene silencing. *Scientific Reports*. 2:509.

36. Lim, C. J., Y. R. Whang, L. J. Kenney, and J. Yan. Gene silencing H-NS paralogue StpA forms a rigid protein filament along DNA that blocks DNA accessibility. *Nucleic Acids Res.* 40:3316-3328.

37. Spurio, R., M. Falconi, A. Brandi, C. L. Pon, and C. O. Gualerzi. 1997. The oligomeric structure of nucleoid protein H-NS is necessary for recognition of intrinsically curved DNA and for DNA bending. *EMBO J.* 16:1795–1805.

38 Winardhi, R. S., W. Fu, S. Castang, Y. Li, S. L. Dove, and J. Yan. 2012. Higher order oligomerization is required for H-NS family member MvaT to form gene-silencing nucleoprotein filament. *Nucleic Acids Res.* 40: 8942–8952.

39. Winardhi, R. S., J. Yan, and L. J. Kenney. 2015. H-NS regulates gene expression and compacts the nucleoid: Insights from single-molecule experiments. *Biophys. J.* 109:1321–1329.

40. Yamanaka, Y., R. S. Winardhi, E. Yamauchi, S. Nishiyama, Y. Sowa, J. Yan, I. Kawagishi, A. Ishihama, and K. Yamamoto. 2018. Dimerization site 2 of the bacterial DNA-binding protein H-NS is required for gene silencing and stiffened nucleoprotein filament formation. *J. Biol. Chem.* 293:9496–9505.

41. Ceci, P., S. Cellai, E. Falvo, C. Rivetti, G. L. Rossi, and E. Chiancone. 2004. DNA condensation and self-aggregation of *Escherichia coli* Dps are coupled phenomena related to the properties of the N-terminus. *Nucleic Acids Res.* 32:5935-5944.

42. Cosgriff, S., K. Chintakayala, Y. T. A. Chim, X. Chen, S. Allen, A. L. Lovering, and D. C. Grainger. 2010. Dimerization and DNA-dependent aggregation of the *Escherichia coli* nucleoid protein and chaperone CbpA. *Mol. Microbiol.* 77:1289-1300.

43. Dadinova, L. A., Y. M. Chesnokov, R. A. Kamyshinsky, I. A. Orlov, M. V. Petoukhov, A. A. Mozhaev, E. Y. Soshinskaya, V. N. Lazarev, V. A. Manuvera, A. S. Orekhov, A. L. Vasiliev, and E. V. Shtykova. 2019. Protective Dps–DNA co-crystallization in stressed cells: an *in vitro* structural study by small-angle X-ray scattering and cryo-electron tomography. *FEBS Letters*. 593:1360–1371

44. Sanchez, A., D. I. Cattoni, J.-C. Walter, J. Rech, A. Parmeggiani, M. Nollmann, and J.-Y. Bouet. 2015. Stochastic self-assembly of ParB proteins builds the bacterial DNA segregation apparatus. *Cell Systems*. 1:163–173.

45. Broedersz, C. P., X. Wang, Y. Meir, J. J. Loparo, D. Z. Rudner, and N. S. Wingreen. 2014. Condensation and localization of the partitioning protein ParB on the bacterial chromosome. *P. Natl. Acad. Sci. USA*. 111: 8809–8814.

46. Funnell, B. E. 2016. ParB partition proteins: Complex formation and spreading at bacterial and plasmid centromeres. *Frontiers in Molecular Biosciences*. 3:44.

47. Graham, T. G.W., X. Wang, D. Song, C. M. Etson, A. M. van Oijen, D. Z. Rudner, and J. J. Loparo. 2014. ParB spreading requires DNA bridging. *Genes Dev.* 28:1228–1238.

48. Amit, R., A. B. Oppenheim, and J. Stavans. 2003. Increased bending rigidity of single DNA molecules by H-NS, a temperature and osmolarity sensor. *Biophys. J.* 84:2467–2473.

49. Hameed, U. F. S., C. Liao, A. K. Radhakrishnan, F. Huser, S. S. Aljedani, X. Zhao, A. A. Momin, F. A. Melo, X. Guo, C. Brooks, Y. Li, X. Cui, X. Gao, J. E. Ladbury, Ł. Jaremko, M. Jaremko, J. Li, and S. T. Arold. 2018. H-NS uses an autoinhibitory conformational switch for environment-controlled gene silencing. *Nucleic Acids Res.* 47:2666–2680.

50. Qin, L., F. Ben Bdira, Y. G. J. Sterckx, A. N. Volkov, J. Vreede, G. Giachin, P. van Schaik, M. Ubbink, and R. T. Dame. 2020. Structural basis for osmotic regulation of the DNA binding properties of H-NS proteins. *Nucleic Acids Res.* 48:2156-2172.

51. Joyeux, M. 2015. Compaction of bacterial genomic DNA: Clarifying the concepts. *J. Phys. Condens. Matter.* 27:383001.

52. Joyeux, M. 2016. *In vivo* compaction dynamics of bacterial DNA: A fingerprint of DNA/RNA demixing ? *Curr. Opin. Colloid Interface Sci.* 26:17-27.

53. Joyeux, M. 2017. Coarse-grained model of the demixing of DNA and non-binding globular macromolecules. *J. Phys. Chem. B.* 121:6351-6358.

54. Joyeux, M. 2018. A segregative phase separation scenario of the formation of the bacterial nucleoid. *Soft Matter*. 14:7368-7381.

55. Florescu, A. M., and M. Joyeux. 2009. Description of non-specific DNA-protein interaction and facilitated diffusion with a dynamical model. *J. Chem. Phys.* 130:015103.

56. Florescu, A. M., and M. Joyeux. 2009. Dynamical model of DNA-protein interaction: effect of protein charge distribution and mechanical properties. *J. Chem. Phys.* 131:105102.

57. Florescu, A. M., and M. Joyeux. 2010. Comparison of kinetic and dynamical models of DNA-protein interaction and facilitated diffusion. *J. Phys. Chem. A.* 114:9662-9672.

58. Jian, H., A. Vologodskii, and T. Schlick. 1997. A combined wormlike-chain and bead model for dynamic simulations of long linear DNA. *J. Comp. Phys.* 136:168-179.

59. Manning, G. S. 1969. Limiting laws and counterion condensation in polyelectrolyte solutions. I. Colligative properties. *J. Chem. Phys.* 51:924-933.

60. Oosawa, F. 1971. Polyelectrolytes. Marcel Dekker, New York.

61. Dorman, C. J., J. C. D. Hinton, and A. Free. 1999. Domain organization and oligomerization among H-NS like nucleoid-associated proteins in bacteria. *Trends Microbiol.* 7:124-128.

62. Link, A. J., K. Robison, and G. M. Church. 1997. Comparing the predicted and observed properties of proteins encoded in the genome of *Escherichia coli* K-12. *Electrophoresis*. 18:1259-1313.

63. Record, M. T., C. F. Anderson, and T. M. Lohman. 1978. Thermodynamic analysis of ion effects on the binding and conformational equilibria of proteins and nucleic acids: the roles of ion association or release, screening, and ion effects on water activity. *Q. Rev. Biophys.* 11:103-178.

64. Mascotti, D. P., and T. M. Lohman. 1990. Thermodynamic extent of counterion release upon binding oligolysines to single-stranded nucleic acids. *P. Natl. Acad. Sci. USA*. 87:3142-3146.

65. Fenley, M. O., C. Russo, and G. S. Manning. 2011. Theoretical assessment of the oligolysine model for ionic interactions in protein-DNA complexes. *J. Phys. Chem. B* 115:9864-9872.

66. Breslauer, K. J., D. P. Remeta, W.-Y. Chou, R. Ferrante, J. Curry, D. Zaunczkowski, J. G. Snyder, and L. A. Marky. 1987. Enthalpy-entropy compensations in drug-DNA binding studies. *P. Natl. Acad. Sci. USA*. 84:8922-8926.

67. Wang, S., A. Kumar, K. Aston, B. Nguyen, J. K. Bashkin, D. W. Boykin, and W. D. Wilson. 2013. Different thermodynamic signatures for DNA minor groove binding with changes in salt concentration and temperature. *Chem. Commun.* 49:8543-8545.

68. Meyer, E. E., K. J. Rosenberg, and J. Israelachvili. 2006. Recent progress in understanding hydrophobic interactions. *P. Natl. Acad. Sci. USA*. 103:15739–15746.

69. Lin, M. S., N. L. Fawzi, and T. Head-Gordon. 2007. Hydrophobic potential of mean force as a solvation function for protein structure prediction. *Structure*. 15:727–740.

70. Makowski, M., C. Czaplewski, A. Liwo, and H. A. Scheraga. 2010. Potential of mean force of association of large hydrophobic particles: Toward the nanoscale limit. *J. Phys. Chem. B*. 114:993–1003.

**FIGURE LEGENDS**

**Figure 1** : Diagrams of protein chains for Models I and II. Index $m$ is indicated for each bead. Red circles represent DNA-binding beads (index $m=1$ and $7$), which rotate freely around beads with index $m=2$ and 6, respectively. Green circles represent isomerization beads (index $m=2$ and 6 for Model I, index $m=2$, 3, 5 and 6 for Model II). In Model I, beads $m=2$ and $m=6$ of one protein chain may bind to beads $m=2$ and $m=6$ of other protein chains. In Model II, beads $m=2$ and $m=6$ of one protein chain may bind to beads $m=3$ and $m=5$ of other protein chains. All other features are common to the two models. Note that the two chains shown in the figure have minimum internal energy.

**Figure 2** : Representative snapshots extracted from simulations with 200 protein chains and $\varepsilon_{LJ} = 11\,k_B T$ for Model I (**left column**) and Model II (**right column**), either without the DNA chain (**top row**) or with the DNA chain (**bottom row**). DNA-binding protein beads are shown in red, isomerization beads are shown in green, other protein beads are not shown. The lines joining the centers of protein beads are shown in black. The line joining the centers of DNA beads is shown in brown (DNA beads are not shown). The blue circle is the trace of the confinement sphere.

**Figure 3** : Plot, as a function of $\varepsilon_{LJ}$, of the average fraction of free (circles), bridging (lozenges), *cis*-bound (triangles), and dangling (upside-down triangles) protein chains for Model I (open symbols) and II (filled symbols). Each set of 4 open or closed symbols with the same value of $\varepsilon_{LJ}$ was obtained from a single simulation with the DNA chain and 200 protein chains, by averaging the relevant quantity over time intervals of at least 2.5 ms after equilibration.

**Figure 4** : Plot, as a function of $\varepsilon_{LJ}$, of the mean radius of the DNA coil **(top)** and the persistence length of the DNA chain **(bottom)**, for Model I (open symbols) and II (filled symbols). In the bottom plot, lozenges represent the values of $\xi_{free}$ and triangles the values of $\xi_{bound}$. The horizontal dot-dashed lines indicate the values of the parameters in the absence of protein chains, that is, $\langle R \rangle = 82.1\,\text{nm}$ and $\xi = 41.7\,\text{nm}$. Each set of open and closed symbols was obtained from a single simulation with the DNA chain and 200 protein chains, by averaging the relevant quantity over time intervals of at least 2.5 ms after equilibration. The error bars in the top plot represent the standard deviation of the fluctuations of $\langle R \rangle$.

**FIGURE 1**

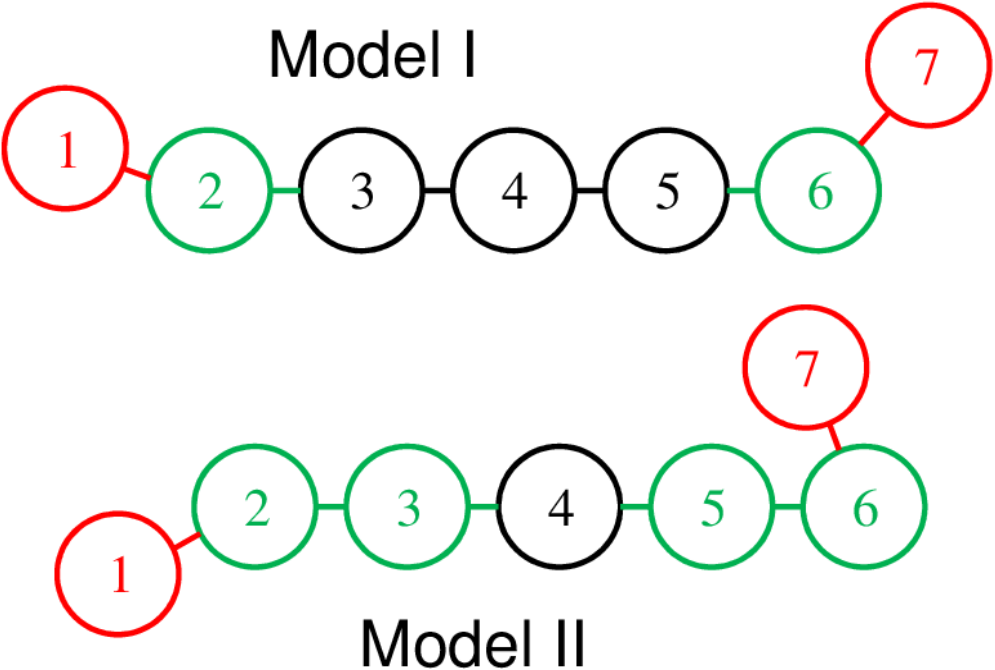

**FIGURE 2**

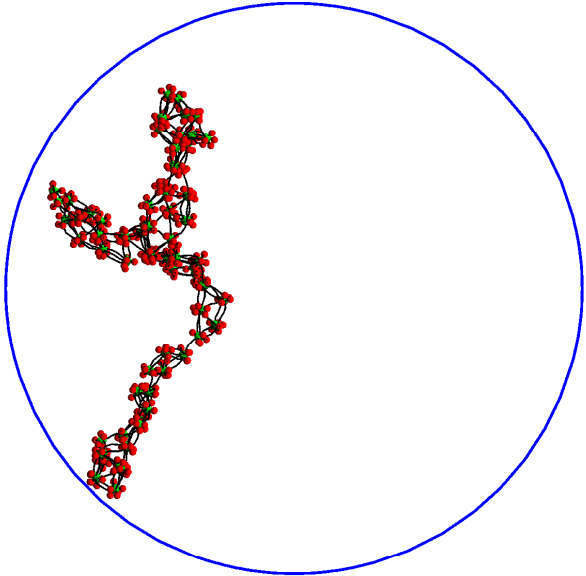
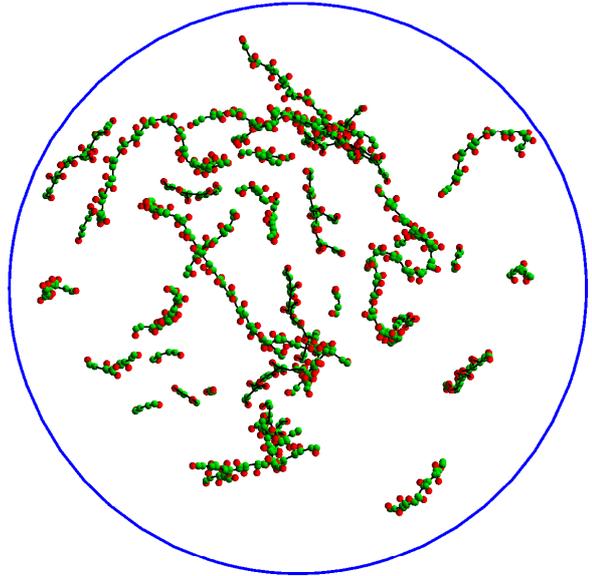
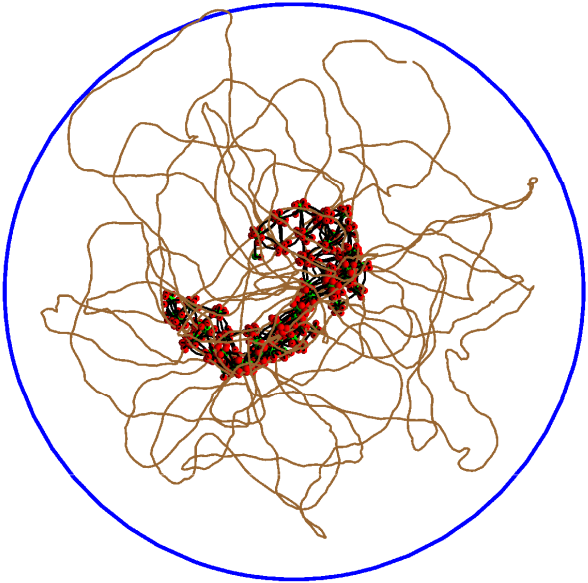
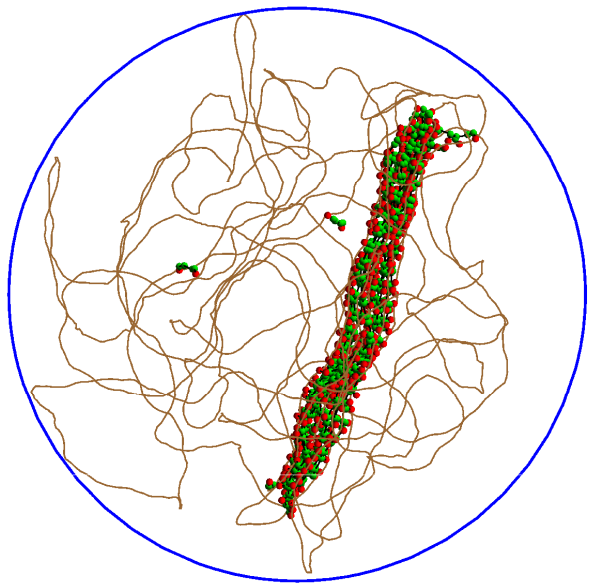

**FIGURE 3**

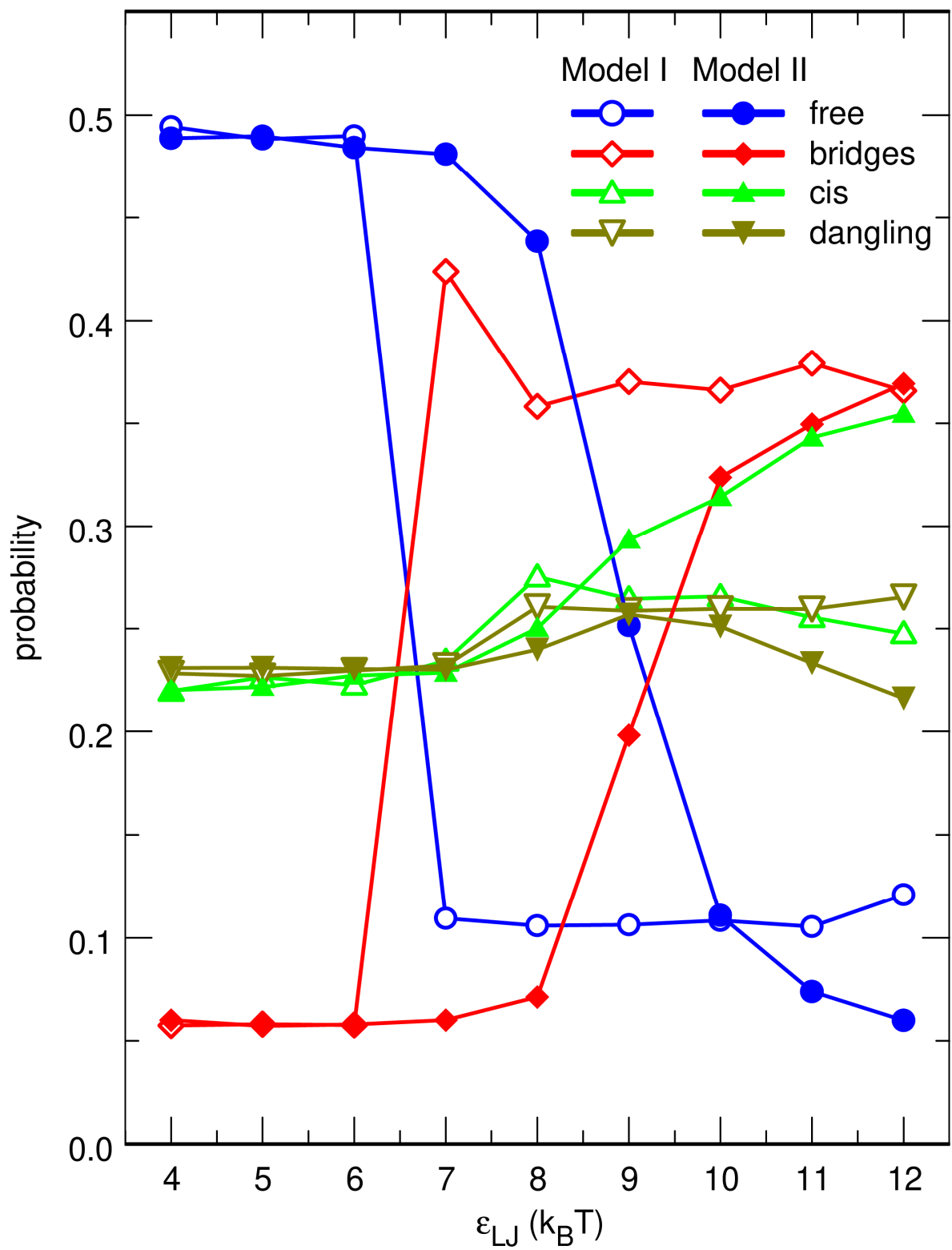

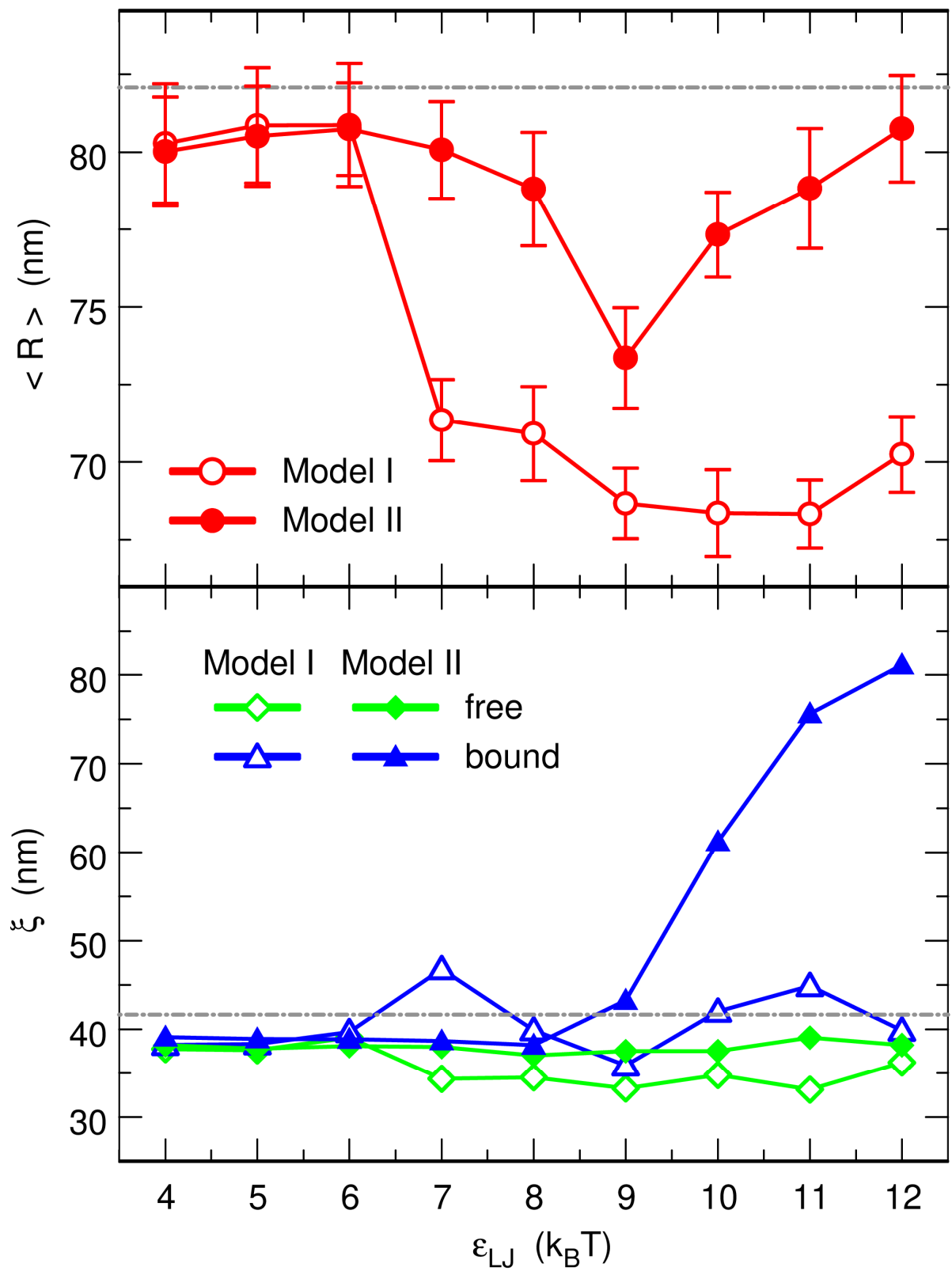

**Impact of self-association on the architectural properties of bacterial nucleoid proteins**

- Supporting Material –

M. Joyeux

*Laboratoire Interdisciplinaire de Physique,*
*CNRS and Université Grenoble Alpes, Grenoble, France*

**MODEL AND SIMULATIONS**

Temperature $T$ is assumed to be 298 K throughout the study. The DNA molecule is modeled as a circular chain of $n = 2880$ beads with radius $a = 1.0$ nm separated at equilibrium by a distance $l_0 = 2.5$ nm and enclosed in a sphere with radius $R_0 = 120$ nm. Each bead represents 7.5 DNA base pairs (bp). The contour length of the DNA molecule and the cell volume correspond approximately to 1/200th of the values for *E. coli* cells, so that the nucleic acid concentration of the model is close to the physiological one (≈10 mM). The potential energy of the DNA chain consists of 4 terms, namely, the stretching energy $V_s^{\text{DNA}}$, the bending energy $V_b^{\text{DNA}}$, the electrostatic repulsion $V_e^{\text{DNA}}$, and a confinement term $V_w^{\text{DNA}}$

$$E^{\text{DNA}} = V_s^{\text{DNA}} + V_b^{\text{DNA}} + V_e^{\text{DNA}} + V_w^{\text{DNA}} \ . \tag{S1}$$

The stretching and bending contributions write

$$V_s^{\text{DNA}} = \frac{h}{2} \sum_{k=1}^{n} (l_k - l_0)^2 \tag{S2}$$

$$V_b^{\text{DNA}} = \frac{g}{2} \sum_{k=1}^{n} \theta_k^2 \ , \tag{S3}$$

where $\mathbf{r}_k$ denotes the position of DNA bead $k$ (with the convention that $\mathbf{r}_{n+k} = \mathbf{r}_k$), $l_k = \|\mathbf{r}_k - \mathbf{r}_{k+1}\|$ the distance between two successive beads, and $\theta_k = \arccos((\mathbf{r}_{k-1} - \mathbf{r}_k)(\mathbf{r}_k - \mathbf{r}_{k+1}) / (\|\mathbf{r}_{k-1} - \mathbf{r}_k\| \|\mathbf{r}_k - \mathbf{r}_{k+1}\|))$ the angle formed by three successive beads. The stretching energy $V_s^{\text{DNA}}$ is a computational device without biological meaning, which is aimed at avoiding a rigid rod description. The stretching force constant $h$ is set to $h = 100 \, k_B T / l_0^2$, which ensures that the variations of the distance between successive beads remain small enough (1). In contrast, the bending rigidity constant is obtained from the known persistence length of the DNA, $\xi = 50$ nm, according to $g = \xi k_B T / l_0 = 20 k_B T$.

Moreover, it is assumed that the repulsion between DNA beads that are not close neighbours along the chain is driven by electrostatics and can be expressed as a sum of repulsive Debye-Hückel potentials with hard core

$$V_e^{DNA} = e_{DNA}^2 \sum_{k=1}^{n-4} \sum_{K=k+4}^{n} H(\|\mathbf{r}_k - \mathbf{r}_K\| - 2a) , \tag{S4}$$

where

$$H(r) = \frac{1}{4\pi\varepsilon r} \exp\left(-\frac{r}{r_D}\right) . \tag{S5}$$

In Eq. (S4), $e_{DNA}$ denotes the electric charge placed at the centre of each DNA bead. The numerical value $e_{DNA} = -3.525\,\overline{e}$, where $\overline{e}$ is the absolute charge of the electron, is the product of $l_0$ and the net linear charge density along a DNA molecule immersed in a buffer with monovalent cations derived from Manning's counterion condensation theory ($-\overline{e}/\ell_B \approx -1.41\,\overline{e}/\text{nm}$) (2,3). In Eq. (S5), $\varepsilon = 80\,\varepsilon_0$ denotes the dielectric constant of the buffer and $r_D = 1.07$ nm the Debye length, whose value corresponds to a concentration of monovalent salt of 100 mM, which includes the (implicit) cationic counterions that are required for the global electroneutrality of the investigated systems. Interactions between close neighbours ($1 \leq |k - K| \leq 3$) are not included in Eq. (S4) because it is considered that they are already accounted for in the stretching and bending terms. The repulsive interaction between two DNA beads is shown as a blue short-dashed line in Fig. S1. Note that the equilibrium distance between two successive beads ($l_0 = 2.5$ nm) is small enough to ensure that different DNA segments do not cross in spite of the small value of $r_D$.

The confinement term $V_w^{DNA}$ is taken as a sum of repulsive terms

$$V_w^{DNA} = 10\, k_B T \sum_{k=1}^{n} f(\|\mathbf{r}_k\|) , \tag{S6}$$

where the function $f$ is defined according to

$$\begin{aligned} f(r) &= 0, \quad \text{if} \quad r \leq R_0 \\ f(r) &= \frac{r^6}{R_0^6} - 1, \quad \text{if} \quad r > R_0 . \end{aligned} \tag{S7}$$

In addition to the DNA chain, the confining sphere contains $P = 200$ DNA-binding protein chains, which corresponds to a protein concentration approximately twice the concentration of H-NS dimers during the cell growth phase and six times the concentration during the stationary phase (4). The number of protein chains was chosen to be as small as possible, but still large enough for the effects discussed in the present paper (compaction and/or stiffening) to be clearly seen in the simulations. $P = 200$ turns out to be an adequate choice in this respect. We note in passing that most experiments dealing with the architectural properties of nucleoid proteins were similarly performed at protein concentrations much larger than physiological ones (5-7). Each DNA-binding protein $j$ is modeled as a chain of 7 beads (indexes $m = 1,2,..,7$), where the two terminal beads $m = 1$ and $m = 7$ represent the DNA-binding sites (see Fig. 1). Protein beads have the same radius $a = 1.0$ nm as DNA beads and are separated at equilibrium by the same distance $l_0 = 2.5$ nm. The internal energy of each protein chain $j$, $E^j$, consists of 4 terms

$$E^j = V_s^j + V_b^j + V_{ev}^j + V_w^j ,$$ (S8)

where the stretching, bending, and confinement contributions are very similar to their DNA counterparts

$$V_s^j = \frac{h}{2} \sum_{m=1}^{6} (L_{jm} - l_0)^2$$ (S9)

$$V_b^j = \frac{g}{2} \sum_{m=3}^{m=5} \Theta_{jm}^2$$ (S10)

$$V_w^j = 10\, k_B T \sum_{m=1}^{7} f(\|\mathbf{R}_{jm}\|) .$$ (S11)

In Eqs. (S9)-(S11), $\mathbf{R}_{jm}$ denotes the position of bead $m$ ($1 \leq m \leq 7$) of protein chain $j$ ($1 \leq j \leq P$), $L_{jm}$ the distance between beads $m$ and $m+1$ of protein chain $j$, and $\Theta_{jm}$ the angle formed by beads $m-1$, $m$, and $m+1$ of protein chain $j$. The values of the stretching force constant $h$ and the bending force constant $g$ are identical to those of the DNA chain. Note, however, that the sum in the expression of $V_b^j$ runs from $m = 3$ to $m = 5$, which means that the DNA-binding beads $m = 1$ and $m = 7$ can rotate without energy penalty around beads $m = 2$ and $m = 6$, respectively. The free rotation of terminal beads mimics the flexible linkers, which connect the C-terminal DNA-binding domains of H-NS to the main body of the dimer (8).

The excluded volume term $V_{ev}^{j}$ ensures that beads belonging to the same chain $j$ repel each other at short distances and do not overlap. $V_{ev}^{j}$ is expressed in terms of the repulsive part of a Lennard-Jones 3-6 potential

$$V_{ev}^{j} = 8\, k_B T \sum_{m=1}^{5} \sum_{M=m+2}^{7} F(\|\mathbf{R}_{jm} - \mathbf{R}_{jM}\| \,|\, r_0) \;, \tag{S12}$$

where $r_0 = 3$ nm and $F(r\,|\,r_0)$ is defined according to

$$F(r\,|\,r_0) = \frac{r_0^6}{r^6} - 2\frac{r_0^3}{r^3} + 1, \quad \text{if} \quad r \leq r_0$$
$$F(r\,|\,r_0) = 0, \quad \text{if} \quad r > r_0. \tag{S13}$$

The repulsive interaction between two protein beads belonging to the same chain is shown as a thin red solid line in Fig. S1.

Interactions between the DNA chain and protein chain $j$, $E^{\text{DNA}/j}$, are taken as the sum of (attractive or repulsive) Debye-Hückel potentials with hard core, which are complemented with repulsive excluded volume terms for DNA-binding beads $m=1$ and $m=7$

$$E^{\text{DNA}/j} = V_e^{\text{DNA}/j} + V_{ev}^{\text{DNA}/j} \;, \tag{S14}$$

where

$$V_e^{\text{DNA}/j} = e_{\text{DNA}} \sum_{k=1}^{n} \sum_{m=1}^{7} e_m H(\|\mathbf{r}_k - \mathbf{R}_{jm}\| - a) \tag{S15}$$

and

$$V_{ev}^{\text{DNA}/j} = 4\, k_B T \sum_{k=1}^{n} \sum_{m=1,7} G(\|\mathbf{r}_k - \mathbf{R}_{jm}\| \,|\, s_0) \;. \tag{S16}$$

Positive charges $e_m = 2.4\,\overline{e}$ are placed at the centre of DNA-binding beads $m=1$ and $m=7$, and negative charges $e_m = -1.2\,\overline{e}$ at the centre of the other beads ($2 \leq m \leq 6$). The values of these effective charges are compatible with those obtained from a naive counting of the number of positively and negatively charged residues in published crystallographic structures of H-NS (9), except that H-NS is globally neutral, whereas the protein chains have here a total slightly negative charge of $-1.2\,\overline{e}$, in agreement with the fact that most proteins encoded in the genome of *E. coli* are anionic (10). In Eq. (S16), $s_0 = 2$ nm and $G(r\,|\,s_0)$ is the repulsive part of a Lennard-Jones 1-2 potential

$$G(r\,|\,s_0) = \frac{s_0^2}{r^2} - 2\frac{s_0}{r} + 1, \quad \text{if} \quad r \leq s_0$$
$$G(r\,|\,s_0) = 0, \quad \text{if} \quad r > s_0. \tag{S17}$$

The interaction between DNA-binding protein beads ($m=1$ and $m=7$) and DNA beads is shown as a thick green long-dashed line in Fig. S1 and the repulsive interaction between other protein beads ($2 \leq m \leq 6$) and DNA beads as a thin green long-dashed line. The most stable DNA/protein complex is shown in Fig. S2(a) : each DNA-binding protein bead binds simultaneously to two successive DNA beads with a total binding energy of $-7.8\,k_BT$, which is comparable to experimentally determined values for complexes of DNA and H-NS ($\approx -11.0\,k_BT$ (11)). It is seen in this figure that the terminal protein bead and the two DNA beads overlap to some extent for the most stable DNA/protein complex. The reason is that the radius of the beads $a=1.0$ nm is essentially used to compute the translational diffusion coefficient $D_t = (k_BT)/(6\pi\eta a)$ which governs the Langevin equations (see Eq. (S25) below). Nonetheless, the distance between the centers of a DNA bead and a protein bead can become smaller than $2a$, which reflects the fact that proteins may insert loops in the major or minor groove of the DNA molecule. The sum of the attractive Debye-Hückel potential with hard core at distance $a$ (Eq. (S15)) and the repulsive part of a Lennard-Jones 1-2 potential without hard core (Eq. (S16)) results in a maximum binding energy of $-3.9\,k_BT$ at a center-center distance of 1.60 nm (thick green long-dashed line in Fig. S1), which is indeed smaller than $2a = 2.0$ nm and is responsible for the overlaps in Fig. S2(a).

The fact that DNA/protein interactions are mediated uniquely by effective electrostatic charges placed at the center of each bead is certainly a strong approximation, because it is known that proteins and cationic counterions compete for binding to the DNA, and that binding of a protein is consequently accompanied by the release of counterions in the buffer (12-14). Since the released counterions regain translational entropy, the net energy balance for the binding of ligands to the DNA results from subtle enthalpy-entropy compensations (15,16), a point which is overlooked in the present models. There is however no reason, why this approximation should affect the validity of the results discussed in the main text, because these results depend essentially on the ability of the proteins to bind to the DNA and on the bond energy, not on the nature of their interactions.

The two models discussed in the main text differ only in the expression of the interaction $E^{j/J}$ between two protein chains $j$ and $J$. The two models agree on the fact that each protein chain contains two isomerization sites, but the properties of these sites differ from Model I to Model II.

In Model I, each isomerization site is made up of a single bead, namely bead $m=2$ for one site and bead $m=6$ for the other one (see Fig. 1). The isomerization beads of protein chain $j$ may bind to the isomerization beads of chain $J$ but repel all other beads through an excluded volume term. The other beads of chain $j$ ($m \neq 2$ and $m \neq 6$) repel all the beads of chain $J$ through the same excluded volume term. More explicitly, for Model I the interaction energy between protein chains $j$ and $J$, $E^{j/J}$, writes

$$E^{j/J} = V_{\text{iso}}^{j/J} + V_{\text{ev}}^{j/J} \ , \tag{S18}$$

where

$$V_{\text{iso}}^{j/J} = \varepsilon_{\text{LJ}} \sum_{\{m,M\} \in \mathfrak{I}} W(\|\mathbf{R}_{jm} - \mathbf{R}_{JM}\| \,|\, u_0) \tag{S19}$$

and

$$V_{\text{ev}}^{j/J} = 8 \, k_{\text{B}} T \sum_{\{m,M\} \notin \mathfrak{I}} F(\|\mathbf{R}_{jm} - \mathbf{R}_{JM}\| \,|\, r_0) \ . \tag{S20}$$

In Eqs. (S19) and (S20), $\mathfrak{I}$ stands for the ensemble

$$\mathfrak{I} = \{\{2,2\},\{2,6\},\{6,6\}\} \ . \tag{S21}$$

In Eq. (S19), the condition $\{m,M\} \in \mathfrak{I}$ indicates that the sum applies to pairs of isomerization beads. In contrast, the restriction $\{m,M\} \notin \mathfrak{I}$ in Eq. (S20) indicates that pairs of isomerization beads do not contribute to the sum. Moreover, in Eq. (S19), $W(r|u_0)$ is a Lennard-Jones 3-6 potential

$$W(r|u_0) = \frac{u_0^6}{r^6} - 2\frac{u_0^3}{r^3} \ , \tag{S22}$$

with $u_0 = 1.0$ nm. The interaction potential of two isomerization protein beads is shown for $\varepsilon_{\text{LJ}} = 8 \, k_{\text{B}} T$ as a thick red solid line in Fig. S1. $W(r|u_0)$ is minimum for $r = u_0$, with $W(u_0|u_0) = -1$, so that for Model I the isomerization binding energy is simply $-\varepsilon_{\text{LJ}}$. The excluded volume term in Eq. (S19) is similar to that in Eq. (S12) and the function $F(r|r_0)$ is defined in Eq. (S13).

In Model II, each isomerization site is instead made up of two beads, namely beads $m = 2$ and $m = 3$ for one site and beads $m = 5$ and $m = 6$ for the other one (see Fig. 1). An isomerization site of one protein chain can bind to an isomerization site of another protein chain, but only in a "head-to-tail" fashion. This is obtained by imposing that beads $m = 2$ and $m = 6$ of one chain can bind to beads $m = 3$ and $m = 5$ of the other chain, but repel all other beads. More explicitly, the interaction energy between protein chains $j$ and $J$, $E^{j/J}$, has the same expression for Model II as for Model I, except that the ensemble $\mathfrak{I}$ is defined according to

$$\Im = \{\{2,3\},\{2,5\},\{3,6\},\{5,6\}\} \ . \tag{S23}$$

$\varepsilon_{LJ}$ is the only variable parameter of the model. It was varied from $4\,k_B T$ to $12\,k_B T$, in order to check the influence of the isomerization binding energy on the equilibrium properties of the system. As already mentioned, for Model I the isomerization binding energy is just $-\varepsilon_{LJ}$. For Model II, the isomerization binding energy of two protein chains was estimated by letting two chains frozen in minimum energy conformations ($\Pi$-shaped) slide parallel to each other and computing their interaction energy as a function of the coordinates of the central bead of one chain relative to the central bead of the other one. The $\Pi$-shaped conformation of protein chains was used for the calculation of the isomerization binding energy, because terminal protein beads rotate without energy penalty around beads $m=2$ and $m=6$ (Eq. (S10)) and the 90° angle ensures minimum repulsion between these beads. As shown for $\varepsilon_{LJ} = 8\,k_B T$ in Fig. S2(b), the interaction energy displays a sharp minimum around $x=7.6$ nm and $y=1.1$ nm, whose depth is taken as the isomerization binding energy. As illustrated in Fig. S3, this energy varies almost linearly with $\varepsilon_{LJ}$, from about $-5\,k_B T$ for $\varepsilon_{LJ} = 4\,k_B T$ to about $-21\,k_B T$ for $\varepsilon_{LJ} = 12\,k_B T$. For the sake of comparison, it is reminded that the experimentally determined value of the enthalpy change upon forming a complex between two H-NS dimers is $-10.2\,k_B T$ (17). For Model II, two protein chains cannot bind simultaneously to the same isomerization site of a third protein chain, which ensures that protein chains isomerize in the form of filaments rather than clusters.

As for DNA/protein interactions, it is stressed that modeling protein/protein interactions as sums of Lennard-Jones potentials is a strong approximation, because protein self-association is governed by a variety of factors, including electrostatic interactions, hydrogen bonds, geometric frustrations, and hydrophobic interactions (18-20). There is however again no reason, why this approximation should affect the validity of the results discussed in the main text, because these results depend essentially on the geometry of protein assemblies and on the strength of the bonds, not on the nature of protein/protein interactions.

The total potential energy of the system, $E_{pot}$, is the sum

$$E_{pot} = E^{DNA} + \sum_{j=1}^{P} E^j + \sum_{j=1}^{P} E^{DNA/j} + \sum_{j=1}^{P-1}\sum_{J=j+1}^{P} E^{j/J} \ . \tag{S24}$$

The dynamics of the model was investigated by integrating numerically the Langevin equations of motion with kinetic energy terms neglected. Practically, the updated position vector for each bead (whether DNA or protein), $\mathbf{r}_j^{(n+1)}$, was computed from the current position vector, $\mathbf{r}_j^{(n)}$, according to

$$\mathbf{r}_j^{(n+1)} = \mathbf{r}_j^{(n)} + \frac{D_t \, \Delta t}{k_B T} \mathbf{F}_j^{(n)} + \sqrt{2 \, D_t \, \Delta t} \, \xi^{(n)} \, , \tag{S25}$$

where the translational diffusion coefficient $D_t$ is equal to $(k_B T)/(6\pi\eta a)$ and $\eta = 0.00089$ Pa s is the viscosity of the buffer at $T = 298$ K. $\mathbf{F}_j^{(n)}$ is the vector of inter-particle forces arising from the potential energy $E_{pot}$, $\xi^{(n)}$ a vector of random numbers extracted at each step $n$ from a Gaussian distribution of mean 0 and variance 1, and $\Delta t$ the integration time step, which was set to 1.0 ps. After each integration step, the position of the center of the confining sphere was slightly adjusted so as to coincide with the center of mass of the DNA molecule.

# SUPPORTING REFERENCES

1.	Jian, H., A. Vologodskii, and T. Schlick. 1997. A combined wormlike-chain and bead model for dynamic simulations of long linear DNA. *J. Comp. Phys.* 136**:**168-179

2.	Manning, G. S. 1969. Limiting laws and counterion condensation in polyelectrolyte solutions. I. Colligative properties. *J. Chem. Phys.* 51:924-933.

3.	Oosawa, F. 1971. Polyelectrolytes. Marcel Dekker, New York.

4.	Azam, T. A., A. Iwata, A. Nishimura, S. Ueda, and A. Ishihama. 1999. Growth phase-dependent variation in protein composition of the *Escherichia coli* nucleoid. *J. Bacteriol.* 181:6361-6370.

5.	Dame, R. T. 2005. The role of nucleoid-associated proteins in the organization and compaction of bacterial chromatin. *Mol. Microbiol.* 56:858-870.

6.	Song, D., and J. J. Loparo. 2015. Building bridges within the bacterial chromosome. *Trends Genet.* 31:164-173.

7.	Liu Y. J., H. Chen, L. J. Kenney, and J. Yan. 2010. A divalent switch drives H-NS/DNA-binding conformations between stiffening and bridging modes. *Genes Dev.* 24:339-344.

8.	Dorman, C. J., J. C. D. Hinton, and A. Free. 1999. Domain organization and oligomerization among H-NS like nucleoid-associated proteins in bacteria. *Trends Microbiol.* 7:124-128.

9.	Arold, S. T., P. G. Leonard, G. N. Parkinson, and J. E. Ladbury. 2010. H-NS forms a superhelical protein scaffold for DNA condensation. *P. Natl. Acad. Sci. USA*. 107:15728-15732.

10.	Link, A. J., K. Robison, and G. M. Church. 1997. Comparing the predicted and observed properties of proteins encoded in the genome of *Escherichia coli* K-12. *Electrophoresis*. 18:1259-1313.

11. Ono, S., M. D. Goldberg, T. Olsson, D. Esposito, J. C. D. Hinton, and J. E. Ladbury. 2005. H-NS is a part of a thermally controlled mechanism for bacterial gene regulation. *Biochem. J.* 391:203-213.

12. Record, M. T., C. F. Anderson, and T. M. Lohman. 1978. Thermodynamic analysis of ion effects on the binding and conformational equilibria of proteins and nucleic acids: the roles of ion association or release, screening, and ion effects on water activity. *Q. Rev. Biophys.* 11:103-178.

13. Mascotti, D. P., and T. M. Lohman. 1990. Thermodynamic extent of counterion release upon binding oligolysines to single-stranded nucleic acids. *P. Natl. Acad. Sci. USA*. 87:3142-3146.

14. Fenley, M. O., C. Russo, and G. S. Manning. 2011. Theoretical assessment of the oligolysine model for ionic interactions in protein-DNA complexes. *J. Phys. Chem. B* 115:9864-9872.

15. Breslauer, K. J., D. P. Remeta, W.-Y. Chou, R. Ferrante, J. Curry, D. Zaunczkowski, J. G. Snyder, and L. A. Marky. 1987. Enthalpy-entropy compensations in drug-DNA binding studies. *P. Natl. Acad. Sci. USA*. 84:8922-8926.

16. Wang, S., A. Kumar, K. Aston, B. Nguyen, J. K. Bashkin, D. W. Boykin, and W. D. Wilson. 2013. Different thermodynamic signatures for DNA minor groove binding with changes in salt concentration and temperature. *Chem. Commun.* 49:8543-8545.

17. Ceschini, S., G. Lupidi, M. Coletta, C. L. Pon, E. Fioretti, and M. Angeletti. 2000. Multimeric self-assembly equilibria involving the histone-like protein H-NS. A thermodynamic study. *J. Biol. Chem.* 275:729-734.

18. Meyer, E. E., K. J. Rosenberg, and J. Israelachvili. 2006. Recent progress in understanding hydrophobic interactions. *P. Natl. Acad. Sci. USA*. 103:15739–15746.

19. Lin, M. S., N. L. Fawzi, and T. Head-Gordon. 2007. Hydrophobic potential of mean force as a solvation function for protein structure prediction. *Structure*. 15:727–740.

20. Makowski, M., C. Czaplewski, A. Liwo, and H. A. Scheraga. 2010. Potential of mean force of association of large hydrophobic particles: Toward the nanoscale limit. *J. Phys. Chem. B*. 114:993–1003.

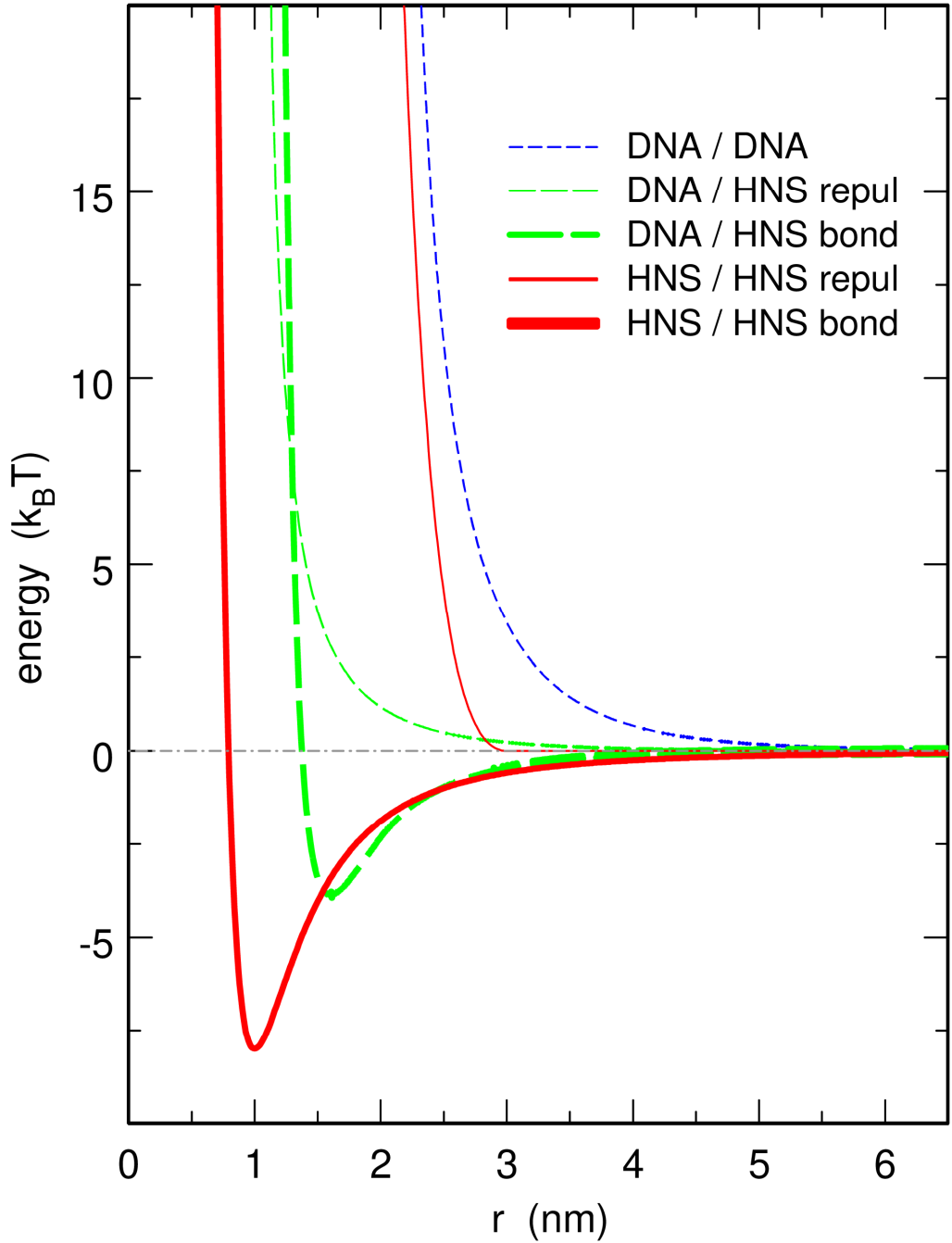

**Figure S1** : Plot, as a function of the distance $r$ between the centers of the beads, of the various bead-bead interaction potentials of the model. ***Blue short-dashed line*** : DNA-DNA repulsive interaction (Eqs. (S4)-(S5)). ***Thin green long-dashed line*** : repulsive potential between a DNA bead and a protein bead with $2 \leq m \leq 6$ (Eq. (S15)). ***Thick green long-dashed line*** : binding potential between a DNA bead and a DNA-binding protein bead with $m=1$ or $m=7$ (Eqs. (S15)-(S17)). ***Thin red solid line*** : repulsive potential between two protein beads, which do not belong to the ensemble $\mathfrak{I}$ (Eqs. (S20), (S21) and (S23)). ***Thick red solid line*** : binding potential between two protein beads, which belong to the ensemble $\mathfrak{I}$ (Eqs. (S19), (S21), (S22) and (S23)). $r$ is expressed in nm, energy values in units of $k_B T$.

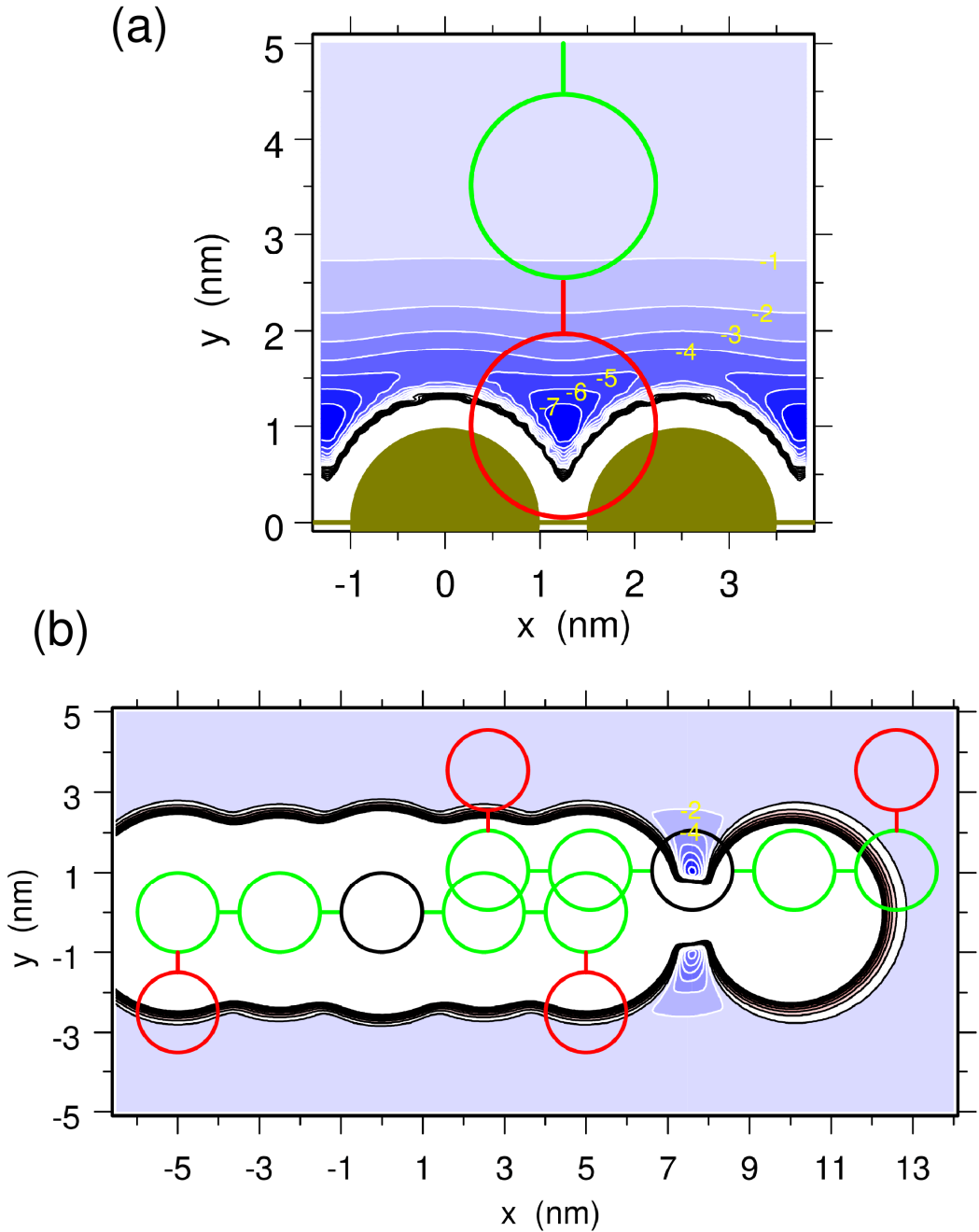

**Figure S2**: **(a)** Diagram showing the most stable DNA-protein complex and how energy evolves when the protein is displaced from this position. Filled brown disks represent DNA beads and circles represent protein beads $m=1$ (red) and $m=2$ (green). Contour lines are separated by $1\,k_BT$. Energy values expressed in units of $k_BT$ are shown in yellow for a few contours. Binding energy is $-7.8\,k_BT$. **(b)** Diagram showing the most stable complex formed by two proteins for Model II and $\varepsilon_{LJ}=8\,k_BT$, and how energy evolves when the upper chain is displaced from this position. Contour lines are separated by $2\,k_BT$. Energy values expressed in units of $k_BT$ are shown in yellow for two contours.. Binding energy is $-12.7\,k_BT$. Contour lines display rotational symmetry around the $x$ axis (the long axis of the fixed protein chain) and point symmetry with respect to the origin of the plot (the center of the central bead of the fixed protein chain).

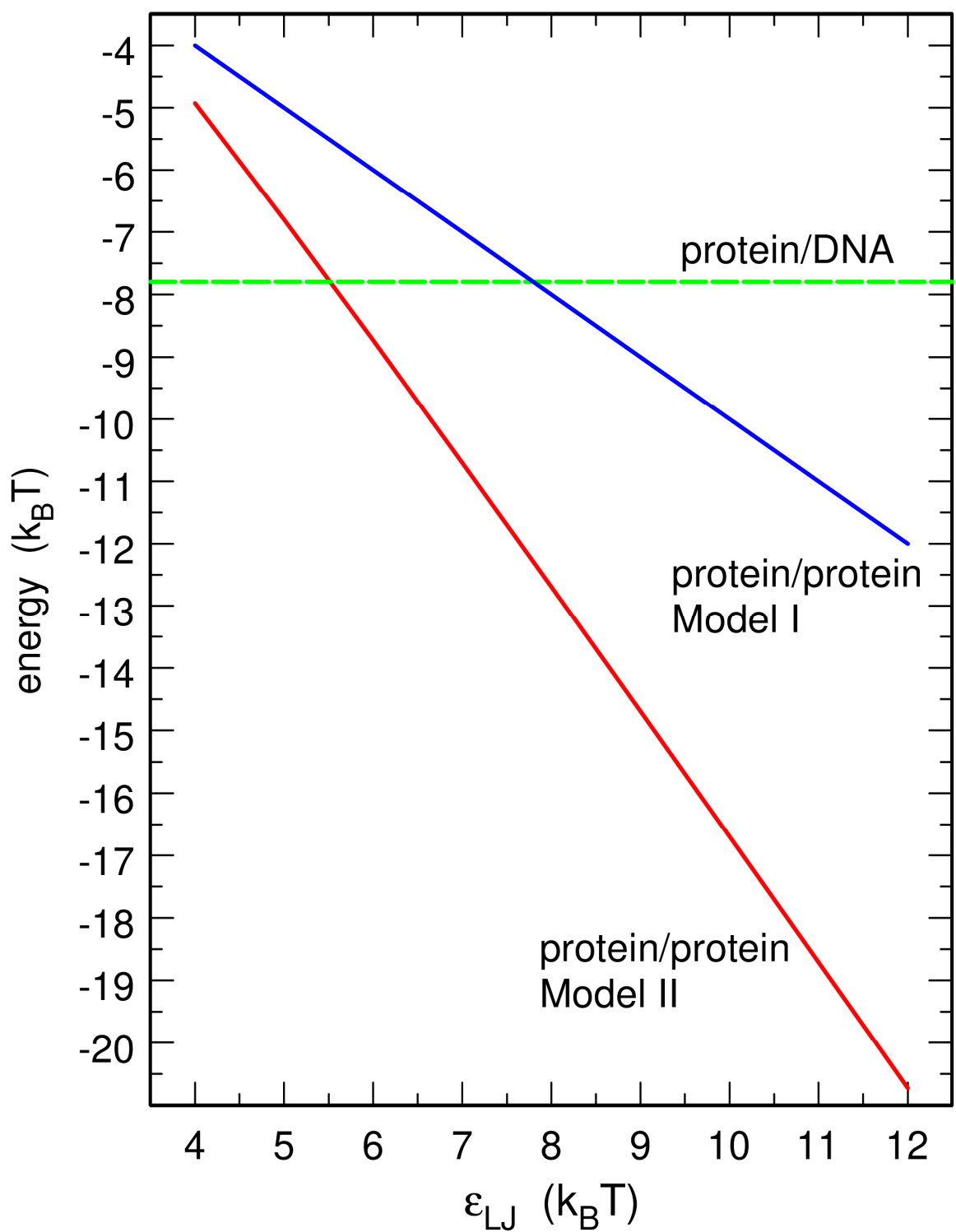

**Figure S3** : The blue and red lines show the evolution, as a function of $\varepsilon_{LJ}$, of the binding energy of protein/protein complexes for Models I and II, respectively. For comparison, the green dashed line shows the bond energy of DNA/protein complexes ($-7.8\,k_B T$). $\varepsilon_{LJ}$ and energy values are expressed in units of $k_B T$.

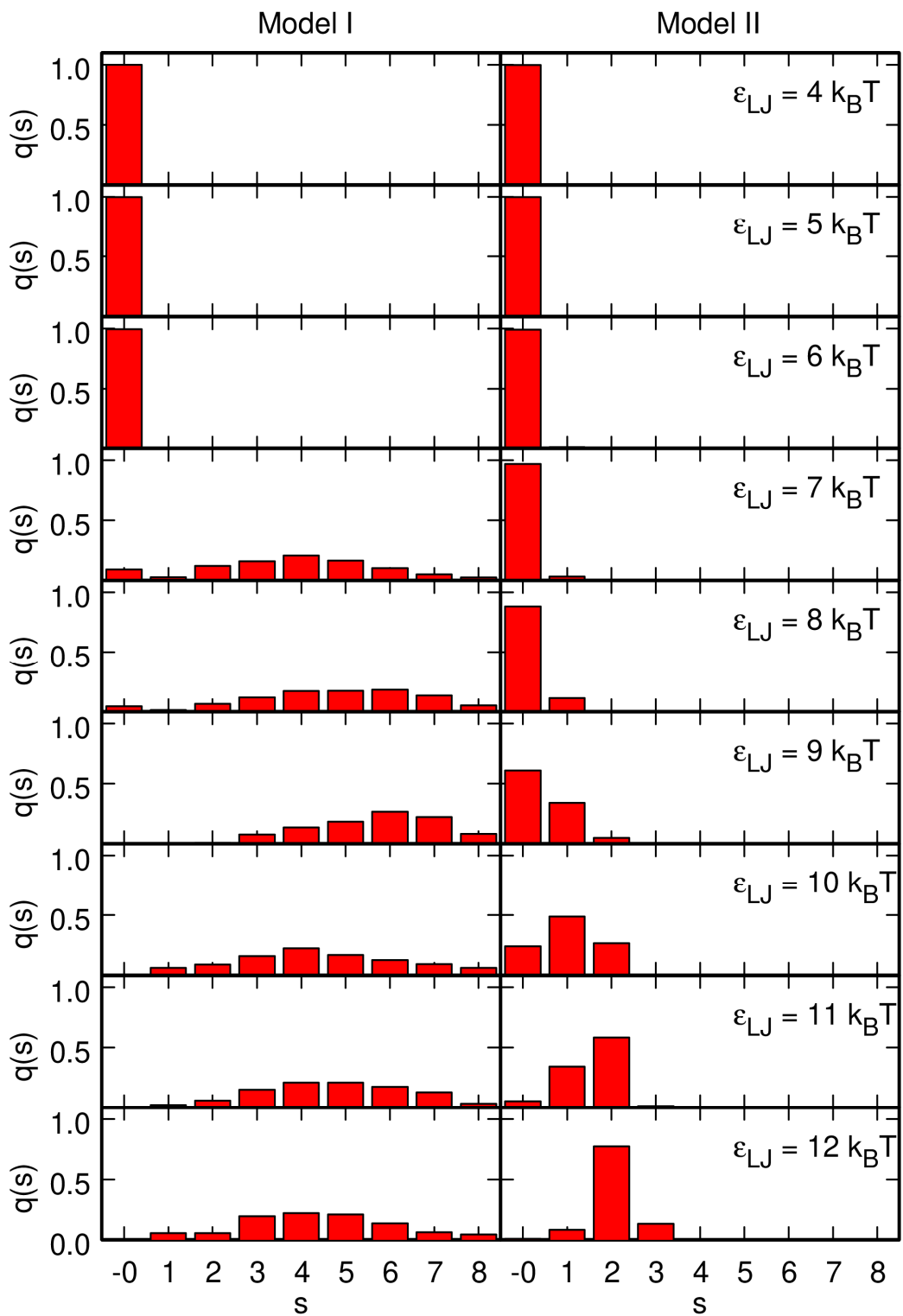

**Figure S4** : Plot of the probability distribution $q(s)$ for a protein chain to bind to $s$ other protein chains, for Model I **(left column)** and Model II **(right column)**, and values of $\varepsilon_{LJ}$ increasing from $4 k_B T$ **(top)** to $12 k_B T$ **(bottom)**. Each plot was obtained from a single simulation with 200 protein chains (without the DNA chain), by averaging $q(s)$ over time intervals of at least 4 ms after equilibration.

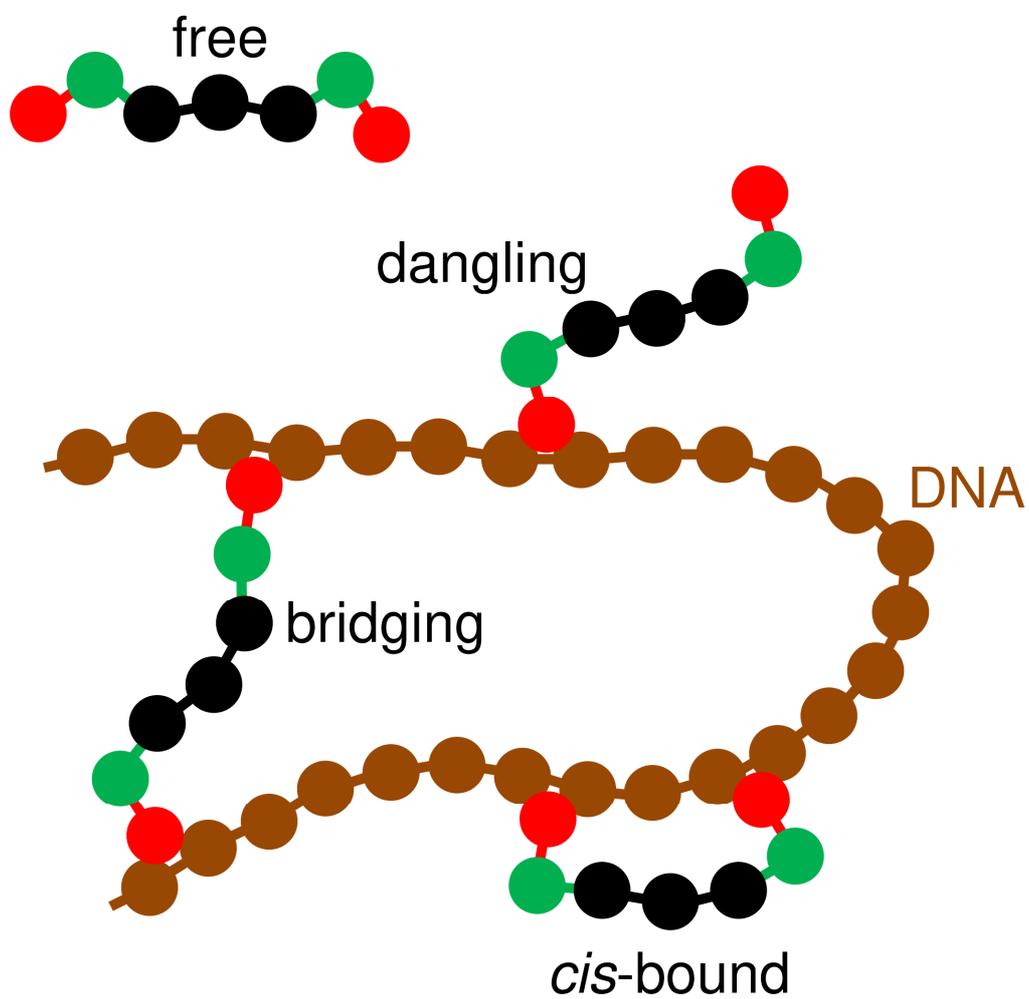

**Figure S5 :** Diagram showing the different modes of interaction between protein chains and a DNA chain. DNA beads are shown in brown and protein beads in red, green or black, according to the same color code as in Fig. 1.

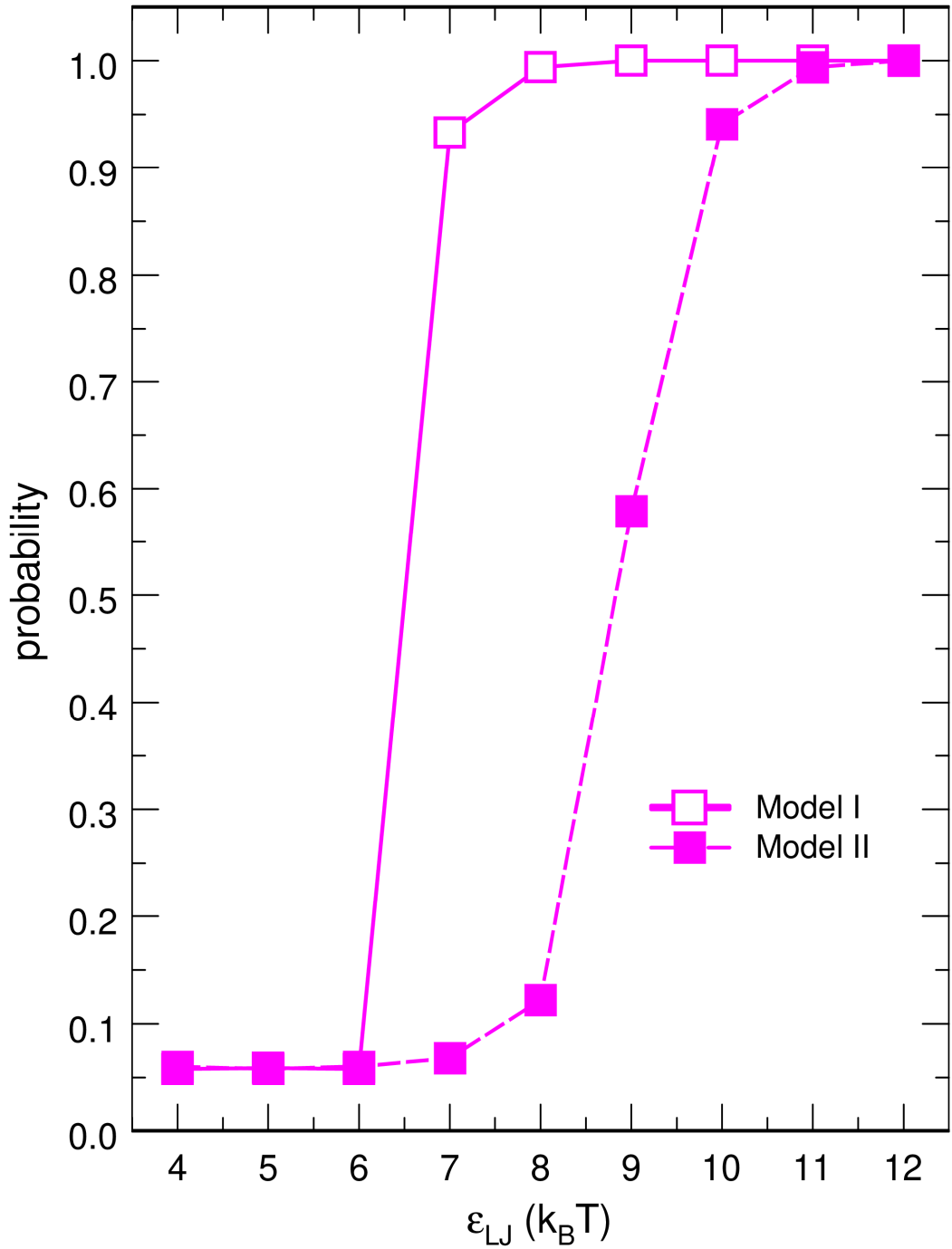

**Figure S6** : Plot, as a function of $\varepsilon_{LJ}$, of the average fraction of protein chains which belong to a cluster bridging different DNA segments, for Model I (open symbols) and II (filled symbols). Each point was obtained from a single simulation with the DNA chain and 200 protein chains, by averaging the relevant quantity over time intervals of at least 2.5 ms after equilibration.

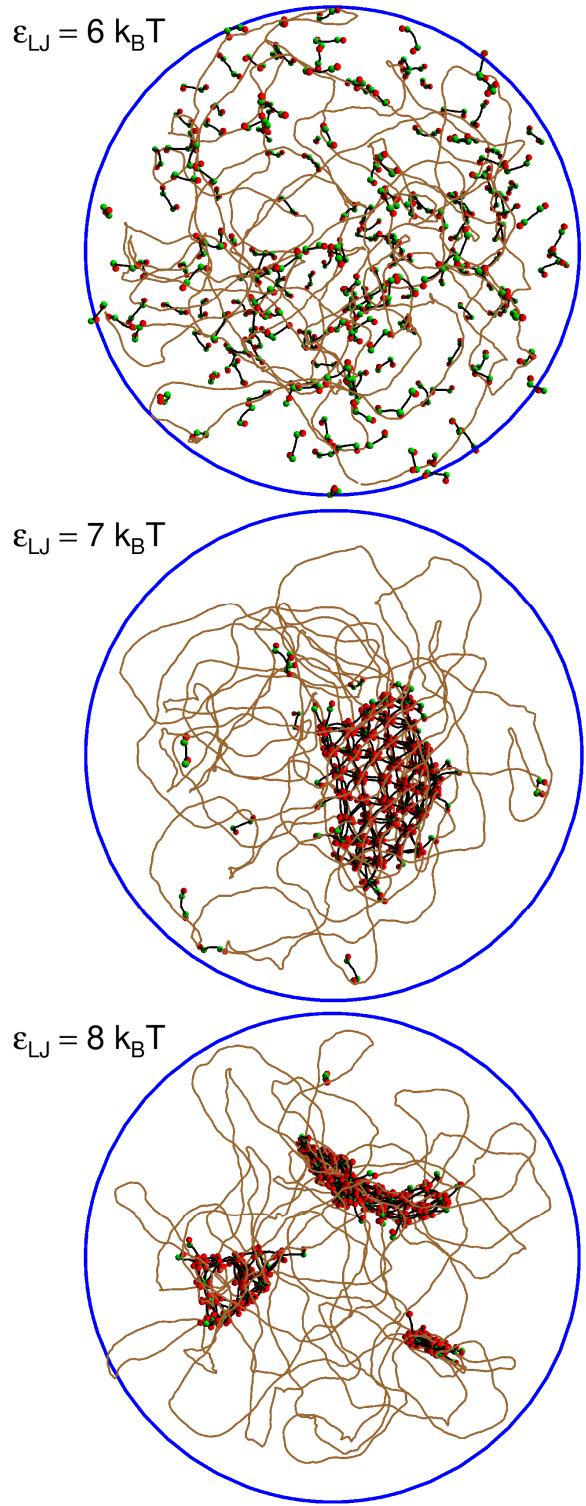

**Figure S7 :** Representative snapshots extracted from simulations with the DNA chain and 200 protein chains for Model I and $\varepsilon_{LJ} = 6\,k_BT$ **(top)**, $7\,k_BT$ **(middle)** and $8\,k_BT$ **(bottom)**. The line joining the centers of DNA beads is shown in brown (DNA beads are not shown). DNA-binding protein beads are shown in red, isomerization beads are shown in green, other protein beads are not shown. The lines joining the centers of protein beads are shown in black. The blue circle is the trace of the confinement sphere.

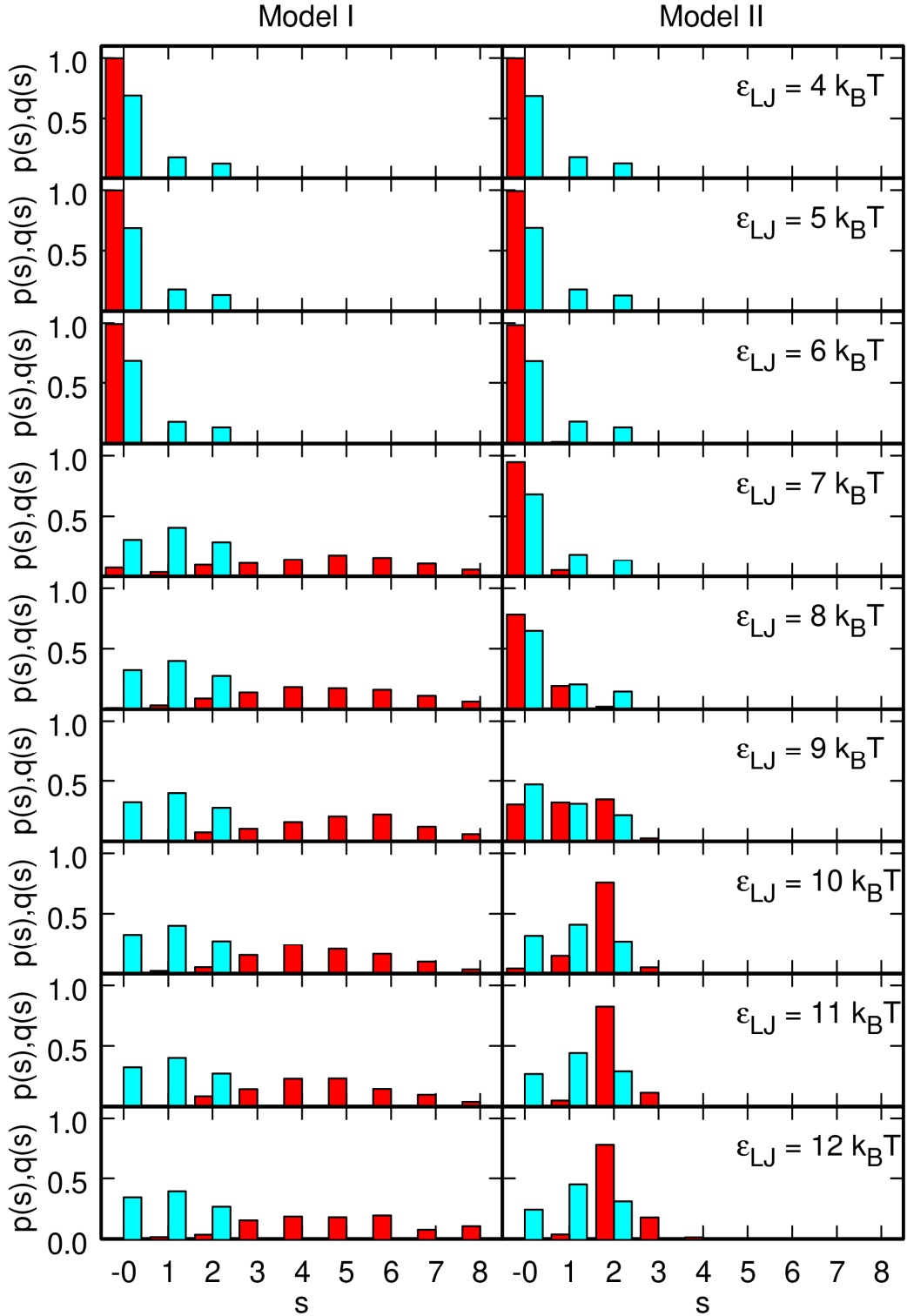

**Figure S8** : Plot of the probability distribution $p(s)$ for a DNA-binding protein bead to bind to $s$ DNA beads **(cyan)** and the probability distribution $q(s)$ for a protein chain to bind to $s$ other protein chains **(red)**, for Model I **(left column)** and Model II **(right column),** and values of $\varepsilon_{LJ}$ increasing from $4k_BT$ **(top)** to $12k_BT$ **(bottom)**. Each plot was obtained from a single simulation with the DNA chain and 200 protein chains, by averaging $p(s)$ and $q(s)$ over time intervals of at least 2.5 ms after equilibration.

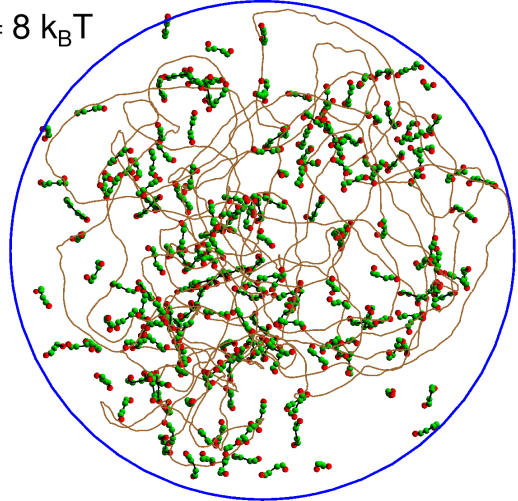

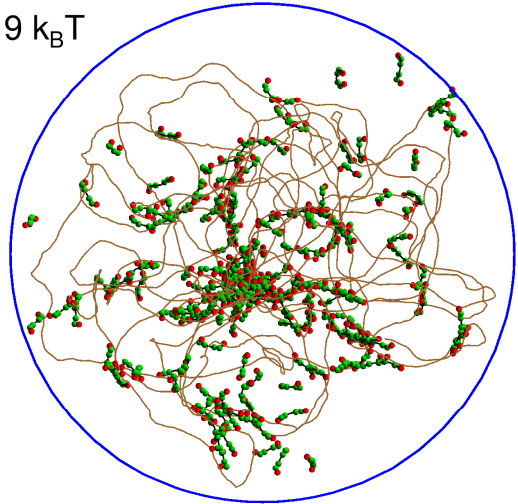

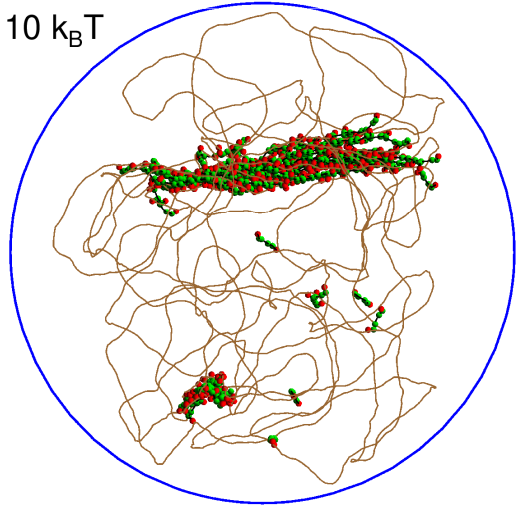

**Figure S9** : Representative snapshots extracted from simulations with the DNA chain and 200 protein chains for Model II and $\varepsilon_{LJ} = 8\,k_BT$ **(top)**, $9\,k_BT$ **(middle)** and $10\,k_BT$ **(bottom)**. The line joining the centers of DNA beads is shown in brown (DNA beads are not shown). DNA-binding protein beads are shown in red, isomerization beads are shown in green, other protein beads are not shown. The lines joining the centers of protein beads are shown in black. The blue circle is the trace of the confinement sphere.